# Understanding Road Usage Patterns in Urban Areas

Pu Wang[1,2], Timothy Hunter[4], Alexandre M. Bayen[4,5], Katja Schechtner[6,7] and Marta C. González[2,3*]

[1]School of Traffic and Transportation Engineering, Central South University, Changsha, Hunan, 410000, P.R. China

[2]Department of Civil and Environmental Engineering, Massachusetts Institute of Technology, Cambridge, MA, 02139, USA

[3]Engineering Systems Division, Massachusetts Institute of Technology, Cambridge, MA, 02139, USA

[4]Department of Electrical Engineering and Computer Sciences, University of California, Berkeley, Berkeley, CA, 94720, USA

[5]Department of Civil and Environmental Engineering, University of California, Berkeley, Berkeley, CA, 94720, USA

[6]Austrian Institute of Technology, Vienna, Vienna, 1210, Austria

[7]MIT Media Lab, Massachusetts Institute of Technology, Cambridge, MA, 02139, USA

**In this paper, we combine the most complete record of daily mobility, based on large-scale mobile phone data, with detailed Geographic Information System (GIS) data, uncovering previously hidden patterns in urban road usage. We find that the major usage of each road segment can be traced to its own - surprisingly few - driver sources. Based on this finding we propose a** *network of road usage* **by defining a bipartite network framework, demonstrating that in contrast to traditional approaches, which define road importance solely by topological measures, the role of a road segment depends on both: its betweeness and its degree in the road usage network. Moreover, our ability to pinpoint the few driver sources contributing to the major traffic flow allows us to create a strategy that achieves a significant reduction of the travel time across the entire road system, compared to a benchmark approach.**

In an era of unprecedented global urbanization, society faces a rapidly accelerating demand for mobility, placing immense pressure on urban road networks[1,2]. This demand manifests in the form of severe traffic congestion[3,4], which decreases the roads' level of service, while at the same time increasing both fuel consumption[5] and traffic-related air pollution[6]. In 2007 alone, congestion forced Americans living in urban areas to travel 4.2 billion hours more, purchase an additional 2.8 billion gallons of fuel, at a total cost of $87.2 billion[3]. To mitigate congestion in urban roads, urban planning[1], traffic prediction[7-9] and the study of complex networks[10-15] have been widely investigated potential influencing factors. However, without comprehensive knowledge of how roads are used dynamically, these studies are conventionally based on expensive and quickly outdated travel surveys or segmented information on traffic flow and travel time[7-9], which fail to support the researchers with the information needed to cope with modern mobility demand. Up to now our understanding of the origins of the drivers in each road remained limited and not quantitatively solved.

In this work we validate for the first time a methodology, which employs comprehensive mobile phone data to detect patterns of road usage and the origins of the drivers. Thus, providing a basis for better informed transportation planning, including targeted strategies to mitigate congestion[3,4]. We formalize the problem by counting the observed number of individuals moving from one location to another, which we put forward as the transient origin destination (*t*-OD) matrix (Fig. S5, Fig. S11 and Supplementary Information (SI) section II.A). Traditionally, ODs are costly and difficult to obtain, because they are at best based on travel diaries made every few years, which quickly become obsolete and strongly rely on provided reports[7,8]. In contrast, the rapidly increasing penetration rates and massive usage of mobile phones, with towers densely located in urban areas, can provide the most detailed information on daily human mobility[16-20] across a large segment of the population[19-25]. Thus we use three-week-long mobile phone billing records generated by 360,000 San Francisco Bay Area users (6.56% of the population, from one carrier) and 680,000 Boston Area users (19.35% of the population, from several carriers) respectively. This data set is two orders of magnitude larger in terms of population and time of observation than the most recent surveys (Table S1), providing us with a source at an unprecedented scale to generate the distribution of travel demands.

To study the distribution of travel demands over a day we divide it into four periods (Morning: 6am-10am, Noon & Afternoon: 10am-4pm, Evening: 4pm-8pm, Night: 8pm-6am) and cumulate trips over the total observational period. A trip is defined when the same mobile phone user is observed in two distinct zones within one hour (zones are defined by 892 towers' service areas in the San Francisco Bay Area and by 750 census tracts in the Boston Area). In the mobile phone data, a user's location information is lost when he/she does not use his/her phone, but by defining the transient origin and destination with movements within one hour, we can capture the distribution of travel demands. Specifically we calculate the *t*-OD as:

$$t\text{-OD}_{ij} = W \times \frac{F^{vehicle}_{ij}}{\sum_{ij}^{A} F^{all}_{ij}} \tag{1}$$

where $A$ is the number of zones. $W$ is the one-hour total trip production in the studied urban area, a number readily available for most cities. However this number gives no information about the trip distribution between zones, which we can enhance by the information gained via mobile phones. Directly from the mobile phone data we calculate $T_{ij}(n)$, which is the total number of trips that user $n$ made between zone $i$ and zone $j$ during the three weeks of study. Via calibrating $T_{ij}(n)$ for the total population we obtain: $F^{all}_{ij} = \sum_{n=1}^{N_k} T_{ij}(n) \times M(k)$, where $N_k$ is the number of users in zone $k$. The ratio $M$ scales the trips generated by mobile phone users in each zone to the trips generated by the total population living there: $M(k) = N_{pop}(k)/N_{user}(k)$, where $N_{pop}(k)$ and $N_{user}(k)$ are the population and the number of mobile phone users in zone $k$. Furthermore to assign only the fraction of the trips attributed to vehicles, we correct $F^{all}_{ij}$ by the vehicle usage rate, which is a given constant for each zone and therefore obtain $F^{vehicle}_{ij}$ (see SI section II.A).

For each mobile phone user that generated the *t*-OD, we can additionally locate the zone where he or she lives, which we define as the *driver source*. Connecting *t*-ODs with driver sources allows us for the first time to take advantage of mobile phone data sets in order to understand urban road usage. In the following, we present the analysis of the road usage characterization in the morning period as a case study. Results for other time periods are presented in SI (Fig. S19).

**Results**

A road network is defined by the links representing the road segments and the nodes representing the intersections. Using incremental traffic assignment, each trip in the *t*-OD matrix is assigned to the road network[26], providing us with estimated traffic flows (Fig. 1a). The road network in the Bay Area serves a considerable larger number of vehicles per hour (0.73 million) than the one in the Boston (0.54 million). The traffic flow distribution $P(V)$ in each area can be well approximated as the sum of two exponential functions corresponding to two different characteristic volumes of vehicles (Fig. 1a); one is the average traffic flow in their arterial roads ($v_A$) and the other is the average traffic flow in their highways ($v_H$). We measure $P(V) = p_A v_A e^{-V/v_A} + p_H v_H e^{-V/v_H}$ ($R^2$>0.99) with $v_A = 373$ (236) vehicles/hour for arterials and $v_H = 1,493$ (689) vehicles/hour for highways in the Bay Area (Boston numbers within parenthesis, $p_A$ and $p_H$ are the fraction of arterial roads and the fraction of highways). Both road networks have similar number of arterials (~20,000), but the Bay Area with more than double the number of highways than Boston (3,141 highways vs 1,267 in Boston) still receives the double of the average flow in the highways ($v_H$) and a larger average flow in the arterial roads.

The volume of vehicles served by a road depends on two aspects: the first is the functionality of the road according to its ability to be a connector based on its location in the road network (i.e. betweenness centrality) and the second is the inherent travel demand of the travelers in the city. The betweenness centrality $b_c$ of a road segment[27-30] is proportional to the number of shortest paths between all pairs of nodes passing through it: we measured $b_c$ by averaging over each pair of nodes, and following the shortest time to destination. The two road networks, analyzed here, have completely different shapes: the Bay Area is more elongated and connects two sides of a bay, while the Boston Area follows a circular shape (see Fig. 2a). But both have a similar function in the distribution of $b_c$: with a broad term corresponding to the arterial roads and an exponential term to the highways, which is at the tail of larger $b_c$. As Fig. 1b shows, we measure: $P(b_c) = p_A P_A(b_c) + p_H P_H(b_c)$ ($R^2$>0.99), with $P_A(b_c) \sim b_c^{-\alpha_A}$ for arterial roads and $P_H(b_c) \sim e^{-b_c/\beta_H}$ for highways. The highways in the Bay Area have an average $b_c$ of $\beta_H = 2.6 \times 10^{-4}$, whereas a larger $\beta_H = 4.6 \times 10^{-4}$ is found for the Boston Area highways, indicating their different topological structures. Interestingly, despite, the different topologies of the two road networks, the similar shapes of their distribution of traffic flows indicate an inherent mechanism in how people are selecting their routes.

Notice that only when the traffic flow is greater than a road's available capacity, the road is congested; the ratio of these two quantities is called Volume over Capacity ($VOC$) and defines the level of service of a road. Surprisingly, despite the different values in average flows $v$ and average betweeness centrality $\beta$, we find the same distribution of $VOC$ (Fig. 1c) in the two metropolitan areas, which follows an exponential distribution with an average $VOC$ given by $\gamma = 0.28$ ($R^2$>0.98):

$$P(VOC) = \gamma e^{-VOC/\gamma} \qquad (2)$$

The exponential decay of $VOC$ indicates that for both road networks traffic flows on 98% of the road segments are well below their designed road capacities, whereas a few road segments suffer from congestion, having a $VOC > 1$. The similarity between the two $VOC$ distributions shows that in both urbanities drivers experience the same level of service, due to utilizing the existing capacities in the similar way.

The traditional difficulty in gathering ODs at large scales has until now limited the comparison of roads in regard to their attractiveness for different driver sources. To capture the massive sources of daily road usage, for each road segment with $V > 0$, we calculate the fraction of traffic flow generated by each driver source, and rank these sources by their contribution to the traffic flow. Consequently, we define a road segment's major driver sources (MDS) as the top ranked sources that produce 80% of its traffic flow. We next define a bipartite network, which we call the *network of road usage*, formed by the edges connecting each road segment to their MDS. Hence, the degree of a driver source $K_{\text{source}}$ is the number of road segments for which the driver source is a MDS, and the degree of a road segment $K_{\text{road}}$ is the number of MDS that produce the vehicle flow in this road segment. As Fig. 2b shows, the driver source's degree $K_{\text{source}}$ is normally distributed, centered in $<K_{\text{source}}> \sim 1000$ in both Bay Area and Boston Area, implying that drivers from each driver source use a similar number of road segments. In contrast, the road segment's degree $K_{\text{road}}$ follows a log-normal distribution (Fig. 2c), where most of the road segments have a degree centered in $<K_{\text{road}}> \sim 20$. This indicates that the major usage of a road segment can be linked to surprisingly few driver sources. Indeed, only 6-7% of road segments are in the tail of the log-normal linked to a larger number of MDS, ranging from 100 to 300.

In Fig. 2a we show a road segment's degree $K_{\text{road}}$ in the road network maps of the Bay Area and the Boston Area. Since census tracts and mobile phone towers are designed to serve similar number of population (Fig. S2), a road segment's degree $K_{\text{road}}$ quantifies the diversity of the drivers using it. We find that $K_{\text{road}}$ is lowly correlated with traditional measures, such as traffic flow, $VOC$ and betweenness centrality $b_c$ (Fig. S15). For example, in Fig. 2a, *Hickey Blvd* in Daly City and *E*

*Hamilton Ave* in Campbell City have a similar traffic flow $V \sim 400$ (vehicles/hour), however, their degrees in the network of road usage are rather different. Hickey Blvd, only has $K_{\text{road}} = 12$, with MDS distributed nearby, whereas E Hamilton Ave, has $K_{\text{road}} = 51$, with MDS distributed not only in its vicinity, but also in some distant areas as Palo Alto, Santa Cruz, Ben Lomond and Morgan Hill.

As Fig. 2a shows, the road segments in the tail of the log-normal ($K_{\text{road}} > 100$) highlight both the highways and the major business districts in both regions. This again implies that $K_{\text{road}}$ can characterize a road segment's role in a transportation network associated with the usage diversity. To better characterize a road's functionality, we classify roads in four groups according to their $b_c$ and $K_{\text{road}}$ in the transportation network (see Fig. 3 and Fig. S16). We define the *connectors*, as the road segments with the largest 25% of $b_c$ and the *attractors* as the road segments with the largest 25% of $K_{\text{road}}$. The other two groups define the highways in the periphery, or *peripheral connectors*, and the majority of the roads are called *local*, which have both small $b_c$ and $K_{\text{road}}$ (Fig. 3). By combining $b_c$ and $K_{\text{road}}$, a new quality in the understanding of urban road usage patterns can be achieved. Future models of distributed flows in urban road networks will benefit by incorporating those ubiquitous usage patterns.

**Discussion**

This novel framework of defining the roads by their connections to their MDS can trigger numerous applications. As a proof of concept, we present here how these findings can be applied to mitigate congestion. For a road segment, its level of congestion can be measured by the additional travel time $t_e$, defined as the difference between the actual travel time $t_a$ and the free flow travel time $t_f$. The drivers who travel through congested roads experience a significant amount of $t_e$. To pinpoint these drivers, the total $T_e$ per driver source is calculated. In contrast to the similar number of population served by each driver source (Fig. S2), the extra travel time $T_e$ generated by driver sources can be very different, following an exponential distribution $P(T_e) = \tau e^{-T_e/\tau}$ (Fig. 4a). Some driver sources present a $T_e$ 16 times larger than the average. This finding indicates that the major traffic flows in congested roads are generated by very few driver sources, which enables us to target the small number of driver sources affected by this significantly larger $T_e$. For the Bay Area, the top 1.5% driver sources (12 sources) with the largest $T_e$ are selected, for the Boston Area we select the top 2% driver sources (15 sources) (Fig. S17). We then reduce the number of trips from these driver sources by a fraction *f*, ranging from 2.7% to 27% in the Bay Area and from 2.5% to 25% in the Boston Area. The reduced numbers of trips correspond to the *m* total percentage of trips (*m* ranging from 0.1% to 1% for both areas). A benchmark strategy, in which trips are randomly reduced without identifying the driver sources with large $T_e$, is used as reference. Our results indicate that the selective strategy is much more effective in reducing the total additional travel time than the random strategy. In the Bay Area, the total travel time reduction $\delta T$ increases linearly with *m* as $\delta T = k(m - b)$ ($R^2$>0.90). We find that when *m*=1%, $\delta T$ reaches 26,210 minutes, corresponding to a 14% reduction of the total Bay Area additional travel time during a one hour morning commute (triangles in Fig. 4b). However, when a random strategy is used, the corresponding $\delta T$ is only 9,582 minutes, which is almost three times less reduction than that achieved by the selective strategy (squares in Fig. 4b). Even better results are found in the Boston Area: using the selective strategy, when *m*=1%, $\delta T$ reaches 11,762 minutes, corresponding to 18% reduction of the total Boston Area additional travel time during a one hour morning commute (diamonds in Fig. 4b), while the random strategy results only in $\delta T = 1,999$ minutes, which is six times less the reduction of that achieved by the selective strategy (circles in Fig. 4b). The underlying reason for the high efficiency of the selective strategy is intrinsically rooted in the two discoveries described above: first that only few road segments are congested and second that most of those road segments can be associated with few MDS.

Today, as cities are growing at an unparalleled pace, particularly in Asia, South America and Africa, the power of our modeling framework is its ability to dynamically capture the massive sources of daily road usage based solely on mobile phone data and road network data, both of which are readily available in most cities. Thus we validate for the first time an efficient method to estimate road usage patterns at a large scale that has a low cost repeatability compared to conventional travel surveys, allowing us to make new discoveries in road usage patterns. We find that two urban road networks with very different demand in the flows of vehicles and topological structures have the same distribution of volume over capacity ($VOC$) in their roads. This indicates common features in the organization of urban trips, which are well captured by the proposed bipartite network of road usage. Based on our findings, a new quality in the understanding of urban road usage patterns can be achieved by combining the traditional classification method of assessing a road's topological importance in the road network, defined by $b_\text{c}$, with the novel parameter of a road's degree in the network of road usage, defined by $K_\text{road}$. The values of $K_\text{road}$ and $b_\text{c}$ together determine a road's functionality. We find that the major traffic flows in congested roads are created by very few driver sources, which can be addressed by our finding that the major usage of most road segments can be linked to their own surprisingly few driver sources. We show the representation provided by the network of road usage is very powerful to create new applications, enabling cities to tailor targeted strategies to reduce the average daily travel time compared to a benchmark strategy.

**Methods**

**Incremental Traffic Assignment.** The most fundamental method to assign trips to road network is provided by the classic Dijkstra algorithm[31]. Dijkstra's algorithm is a graph search algorithm that solves the shortest path problem for a graph with nonnegative edge path costs (travel time in our case). However, the Dijkstra algorithm ignores the dynamical change of travel time in a road segment. Thus to incorporate the change of travel time, we apply the incremental traffic assignment (ITA) method[26] to assign the *t*-OD pairs to the road networks. In the ITA method, the original *t*-OD is first split into four sub *t*-ODs, which contain 40%, 30%, 20% and 10% of the original *t*-OD pairs respectively. These fractions are the commonly used values[32]. The trips in the first sub *t*-OD are assigned using the free travel time $t_f$ along the routes computed by Dijkstra's algorithm. After the first assignment, the actual travel time $t_a$ in a road segment is assumed to follow the Bureau of Public Roads (BPR) function that widely used in civil engineering $t_a = t_f(1 + \alpha(VOC)^\beta)$, where commonly used values $\alpha = 0.15$ and $\beta = 4$ are selected[32]. Next, the trips in the second sub *t*-OD are assigned using the updated travel time $t_a$ along the routes computed by Dijkstra's algorithm. Iteratively, we assign all of the trips in the four sub *t*-ODs. In the process of finding the path to minimize the travel time, we record the route for each pair of transient origin and transient destination (see SI section II.B for more detail).

**Validating the predicted traffic flow by probe vehicle GPS data.** Due to the lack of reliable traffic flow data at a global scale, we compare for each road segment the predicted travel time with the average travel time calculated from probe vehicle GPS data. According to the BPR function, the travel time of a road segment is decided by its traffic flow: a road segment's travel time increases with the increase of its traffic flow. Hence, obtaining the travel time from GPS probe data is an independent way to validate our result on the distribution of traffic flow. We find a very good linear relation $T_\text{prediction} = kT_\text{probe vehicle}$ with both travel times obtained independently (the coefficient of determination $R^2 > 0.9$ for all time periods, see SI section II.C for more detail).

**Calculation of driver sources.** A driver source is calculated from the mobile phone data based on the regularity of visits of mobile phone users at each time of the day[20]. This regularity is time dependent, and peaks at night when most people tend to be reliably at a home base with an average probability of 90% (Fig. S6B). Thus, we make a reasonable assumption that a driver source is the zone where the user is mostly found from 9pm to 6am in the entire observational period.

## Acknowledgements

The authors thank A.-L. Barabási, H. J. Herrmann, D. Bauer, J. Bolot and M. Murga for valuable discussions. This work was funded by New England UTC Year 23 grant, awards from NEC Corporation Fund, the Solomon Buchsbaum Research Fund and National Natural Science Foundation of China (No. 51208520). P. Wang acknowledges support from Shenghua Scholar Program of Central South University.

## Author contributions

PW and MG designed research; PW, TH, AB and M.G performed research; PW, KS and MG wrote the paper.

## Additional Information

**Competing financial interests:** The authors declare no competing financial interests.

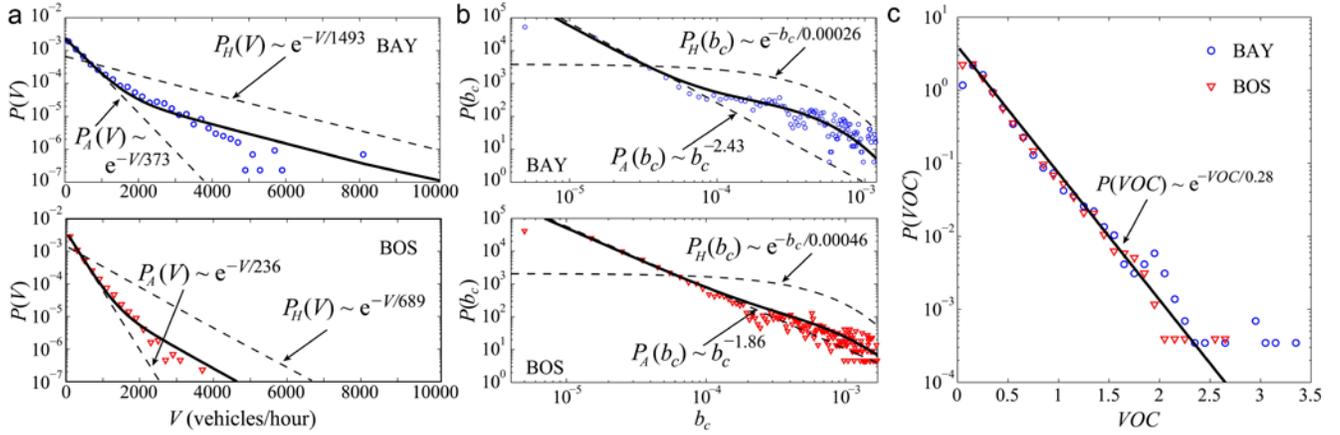

**Figure 1 | Distributions of traffic flow, betweenness centrality and $VOC$ in the two urban areas.** (a) The one-hour traffic flow $V$ follows a mixed exponential distribution $P(V) = p_A v_A e^{-V/v_A} + p_H v_H e^{-V/v_H}$ for both Bay Area and Boston Area, where constants $p_A$ and $p_H$ are the fraction of arterial roads and the fraction of highways, $v_A$ and $v_H$ is the average traffic flow for arterial roads and highways respectively. (b) The distribution of road segment's betweenness centrality $b_c$ is well approximated by $P(b_c) = p_H \beta_H e^{-b_c/\beta_H} + p_A c_A b_c^{-\alpha_A}$, where the power-law distribution approximates arterial roads' $b_c$ distribution and the exponential distribution approximates highways' $b_c$ distribution. $\beta_H$ denotes the average $b_c$ of highways and $\alpha_A$ is the scaling exponent for the power-law. (c) The volume over capacity $VOC$ follows an exponential distribution $P(VOC) = \gamma e^{-VOC/\gamma}$ with an average $VOC = 0.28$ for the two areas. Traffic flows in most road segments are well under their designed capacities, whereas a small number of congested segments are detected. For more statistical analysis of the fits, see the detailed discussion in SI section III.B.

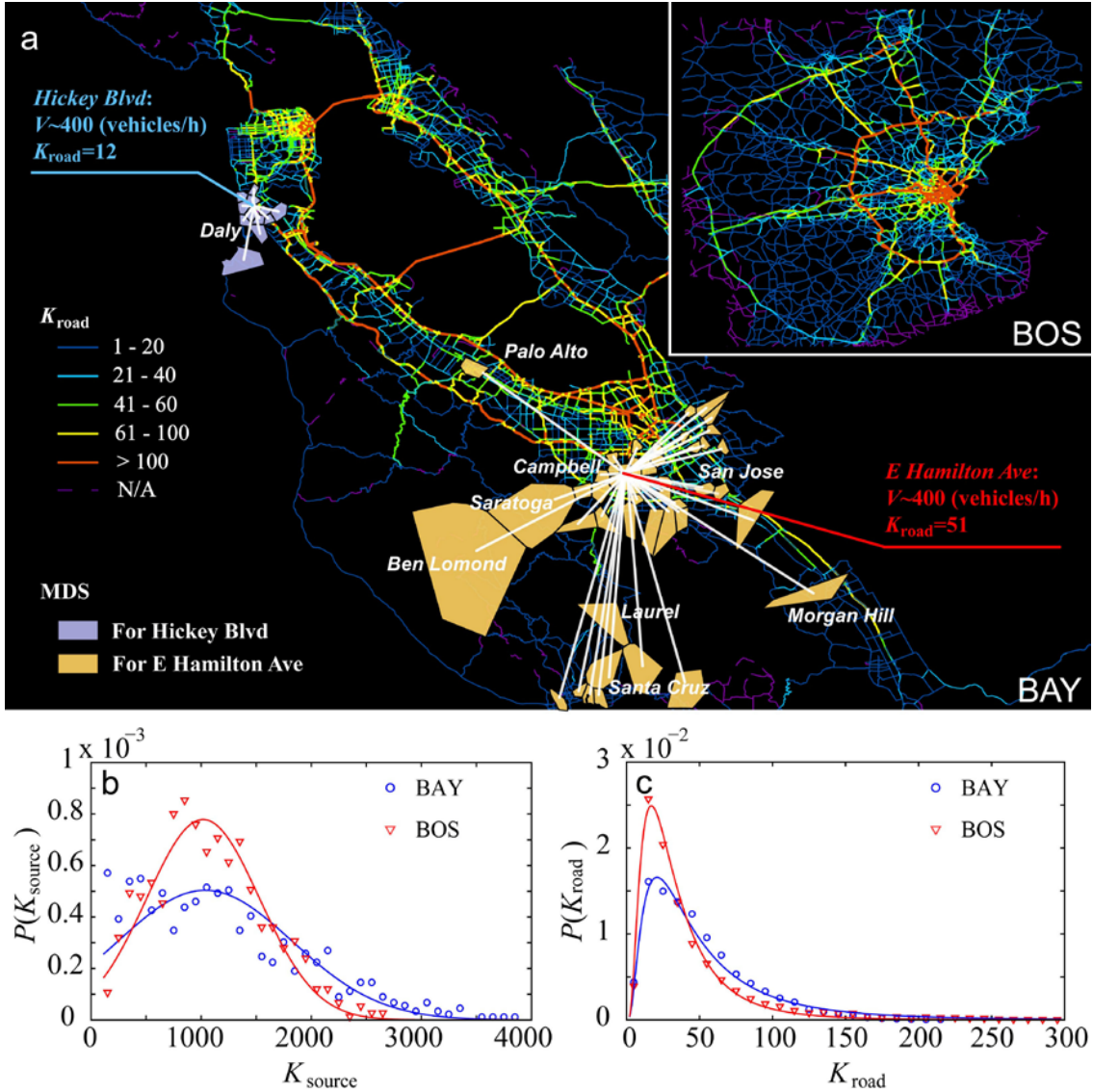

**Figure 2 | Tracing driver sources via the road usage network.** (a) The colour of a road segment represents its degree $K_{road}$. Most residential roads are found to have small $K_{road}$, whereas the backbone highways and the downtown arterial roads are shown to have large $K_{road}$. The light blue polygons and the light orange polygons pinpoint the MDS for *Hickey Blvd* and *E Hamilton Ave* respectively. The white lines show the links that connect the selected road segment and its MDS. The two road segments have a similar traffic flow $V \sim 400$ (vehicles/hour), however, *Hickey Blvd* only has 12 MDS located nearby, whereas *E Hamilton Ave* has 51 MDS, not only located in the vicinity of *Campbell City*, but also located in a few distant regions pinpointed by our methodology. (b) The degree distribution of driver sources can be approximated by a normal distribution $P(K_{source}) = e^{-(K_{source} - \mu_{source})^2 / 2\sigma_{source}^2} / (\sqrt{2\pi}\sigma_{source})$ with $\mu_{source} = 1{,}035.9$ $(1{,}017.7)$, $\sigma_{source} = 792.2$ $(512.3)$, $R^2 = 0.78$ $(0.91)$ for Bay Area (Boston Area). (c) The degree distribution of road segments is approximated by a log-normal distribution $P(K_{road}) = e^{-(\ln(K_{road}) - \mu_{road})^2 / 2\sigma_{road}^2} / (\sqrt{2\pi}\sigma_{road} K_{road})$ with $\mu_{road} = 3.71$ $(3.36)$, $\sigma_{road} = 0.82$ $(0.72)$, $R^2 = 0.98$ $(0.99)$ for Bay Area (Boston Area). For more statistical analysis of the fits, see SI section III.B.

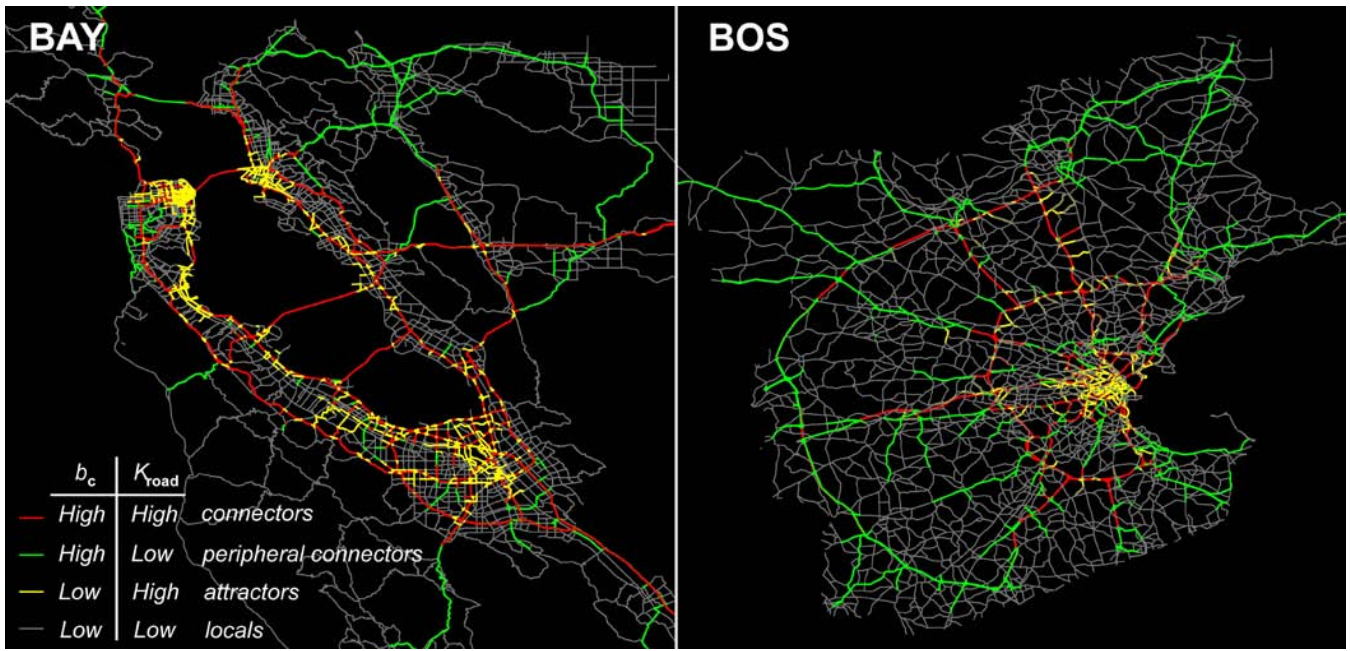

**Figure 3 | Types of roads defined by $b_c$ and $K_{road}$.** The road segments are grouped by their betweenness centrality $b_c$ and degree $K_{road}$. The red lines (connectors) represent the road segments with the top 25% of $b_c$ and $K_{road}$; they are topologically important and diversely used by drivers. The green lines (peripheral connectors) represent the road segments in the top 25% of $b_c$, but with low values of $K_{road}$; they are topologically important, but less diversely used. The road segments in yellow are those with low values of $b_c$, but within the top 25% $K_{road}$; they behave as attractors to drivers from many sources (attractors). The road segments in grey have the low values of $b_c$ and $K_{road}$, they are not topologically important and locally used (locals).

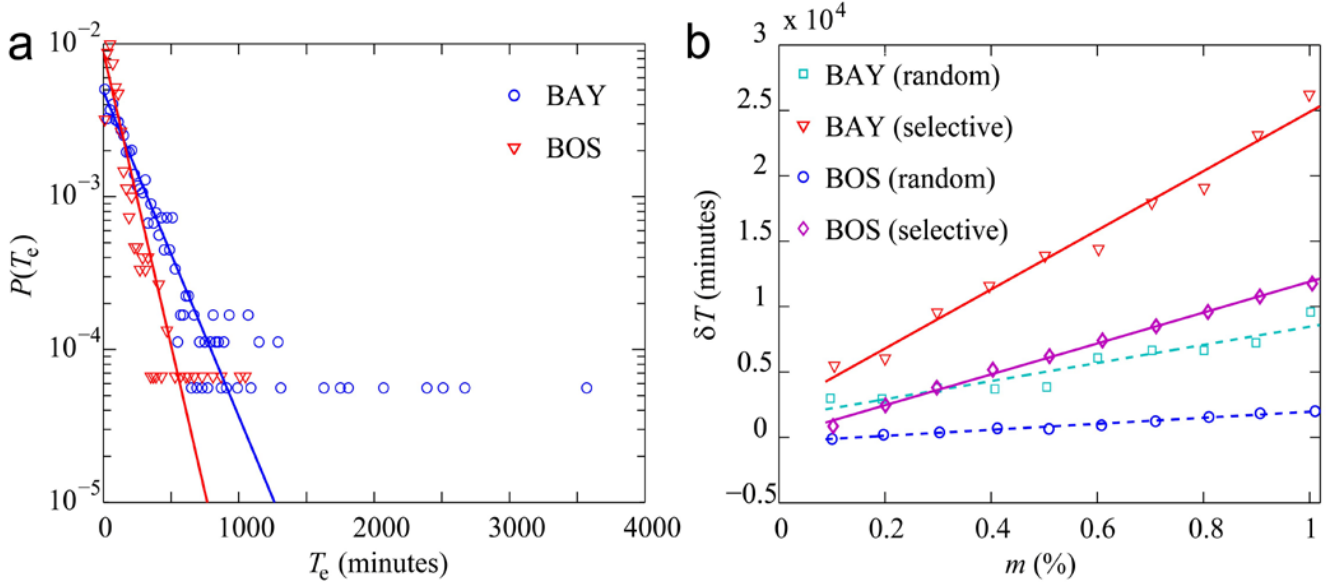

**Figure 4 | A selective strategy based on the framework of road usage network shows better efficiency in mitigating traffic congestion.** (a) The distribution of the additional time $T_e$ for each driver source is well approximated with an exponential distribution $P(T_e) = \tau e^{-T_e/\tau}$. $T_e$ is unevenly distributed in the two urban areas (also see Fig. S17). The maximum $T_e$ (Bay Area 3,562 minutes and Boston Area 1,037 minutes) is significantly larger than the average $T_e$ (204 minutes and 113 minutes respectively). (b) The total additional travel time reduction $\delta T$ according to the trip reduction percentage $m$ for the selective vs. the random strategy. The fits correspond to $\delta T = k(m - b)$, where $k$ is the slope of the linear fit, $b$ is close to zero for all fits. For a detailed statistical analysis of the fits, see the discussion in SI section III.B.

# Supplementary Information

Pu Wang, Timothy Hunter, Alexandre M. Bayen, Katja Schechtner & Marta C. González

## TABLE OF CONTENTS

# I. DATA

## A. Mobile Phone Data and Census Tract Data

This section describes the data used in the main article. To this day these are the most extensive data sets which have been used to perform road usage studies. The San Francisco Bay Area mobile phone data are collected by a US mobile phone operator and contain about half a million customers. Each time a person uses a phone (call/text message/web browsing) the time and the mobile phone tower providing the service is recorded. This altogether generates 374 million location records in the three week observational period. A voronoi tessellation is used to estimate the service area of a mobile phone tower (1, 2). It provides the rough region where a mobile phone user can be located by his/her phone usage (Fig. S1A). The voronoi polygons located at the border are reshaped along the outline border of the San Francisco Bay Area census tracts to guarantee that they have reasonable service areas (Fig. S1A). Among these half a million users, we select 356,670 users to study the travel demands of the Bay Area residents (Table S1).

| Properties: | Bay Area | Boston Area |
|---|---|---|
| Population | 5,434,155 | 3,528,930 |
| Area (mile$^2$) | 3,746 | 1,825 |
| Population Density (/mile$^2$) | 1,451 | 1,934 |
| Avg. Car Pool Size (people per car) | 2.25 | 2.16 |
| Mobile Phone Users | 356,670 | 683,001 |
| Total Length of Road Segments (miles) | 15558.4 | 10346.5 |
| Total Length of Road Segments/Population (miles/person) | 0.00286 | 0.00293 |
| Number of Arterial Roads | 21,267 | 20,638 |
| Number of Highways (Including Freeways) | 3,141 | 1,267 |

**Table S1.** General information extracted from mobile phone data, census tract data and GIS data. The selected mobile phone users represent 6.56% and 19.35% of the population in the two metropolitan areas respectively. This is roughly two orders of magnitude larger in terms of population and time of observation than the most recent surveys (3). The length of road segments takes into account the num of lanes of a road segment.

In the Boston Area the coordinates of the recorded locations are estimated by a standard triangulation algorithm (location data do not come with tower ID). In the three weeks' observational period, more than 200,000 distinct locations are recorded, this data is aggregated at the census tract level to define the location of a phone user (Fig. S1D). Consequently, we select 683,001 users from the one million mobile phone users in the Boston Area.

In both areas the selected mobile phone users have at least one location recorded between 9:00pm to 7:00am, allowing for the definition of home location in connection with a tower's service area or a census tract. The mobile phone users' home locations are also defined as the *driver sources*. We further find that a large majority of driver sources are located within dense mobile phone grids or small enough census tracts, thus providing accurate spatial resolution for the purpose of this study. The area distributions of driver sources are illustrated in Fig. S1B and E, and the respective density of population in Fig. S1C and F.

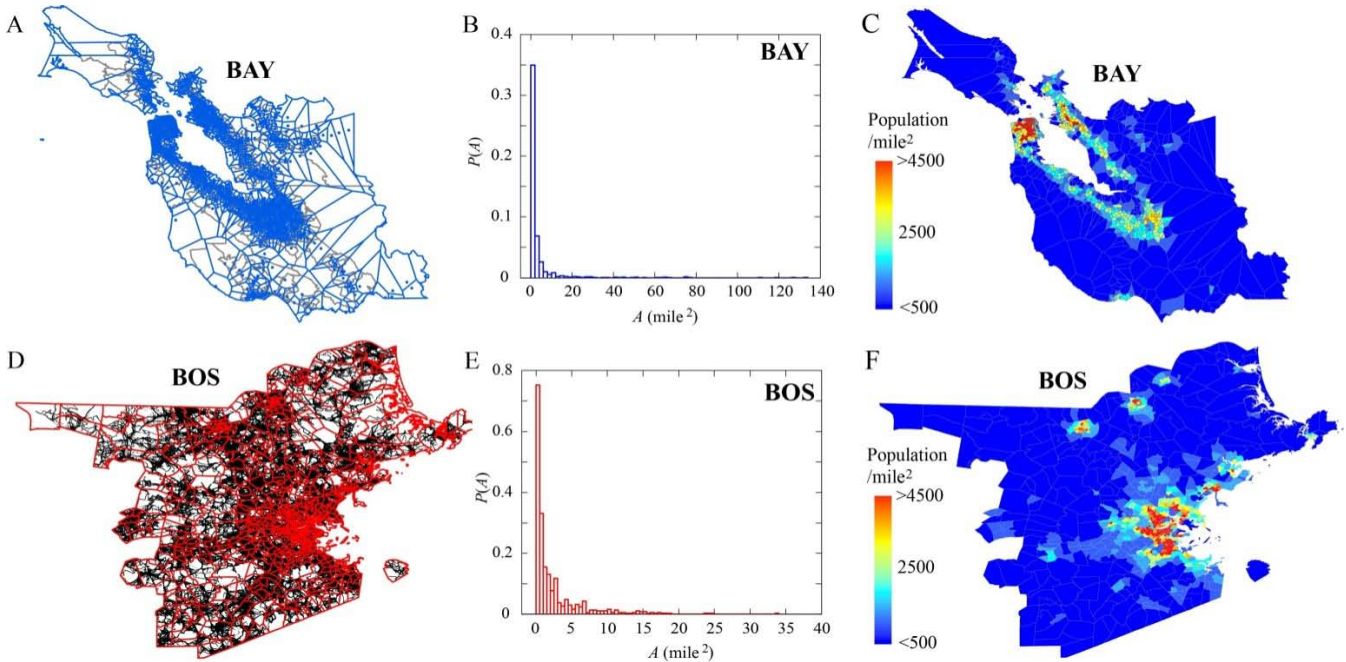

**Figure S1.** Location data and driver sources. (**A**) In the Bay Area (BAY), 892 mobile phone towers (blue dots) are used by the carrier. The covering areas of the towers are defined by a voronoi tessellation (blue polygons). The census tracts are represented by the light grey polygons. (**B**) The area distribution of Bay Area driver sources $P(A)$ quantifies the probability that a driver source has an area $A$. The areas of most driver sources are small, indicating a high accuracy of driver sources' locations. (**C**) In the Bay Area, the population density of each driver source is calculated by the population of its overlapping census tracts. (**D**) In the Boston Area (BOS) driver sources are defined by census tracts (red polygons,

750 in total). Mobile phone users' coordinates are estimated by a standard triangulation algorithm, which results in more than 200,000 distinct locations with a 100m×100m spatial resolution (black dots). (**E**) Same with (B) for the Boston Area. (**F**) The population density in a Boston Area driver source is derived from the census tract data.

As shown in Fig. S2, we measure the population in each driver source. Since mobile phone towers and census tracts are designed to serve similar number of population, we find that diver sources have a similar order of magnitude.

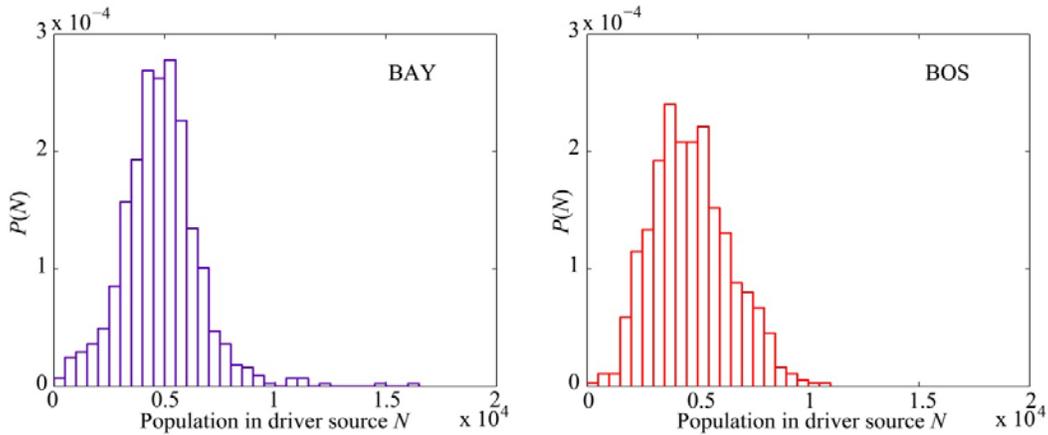

**Figure S2.** The distribution of population in driver sources. *N* is the population of a driver source. In the Bay Area, a driver source is a mobile phone tower's service area. In the Boston Area, a driver source is a census tract.

Users' privacy is protected by using anonymized user IDs. In addition, the spatial resolution of the voronoi lattice or the census tract provides sufficiently large areas to prevent personal location identification at an individual level. Furthermore, no individual trajectory is shown in our results.

## B. Road Network Data

The road networks, which include both highways and arterial roads, are provided by NAVTEQ, a commercial provider of geographical information systems data (4). The data incorporate the attributes of roads needed for the computations presented in this work, in particular the road capacity. The road network in the Bay Area contains 21,880 road segments and 11,096 intersections, while the road network in the Boston Area contains 21,905 road segments and 9,643 intersections. For each road segment, the speed limit $sl$ (miles/hr), the number of lanes $l$ and the direction are extracted from the database. According to 2000 Highway Capacity Manual (5) and Reference (6), we estimate the capacity $C$ of a road segment as follows:

(1) when the speed limit of a road segment $sl \leq 45$, it is defined as an arterial road:

$$C=1,900 \times l \times q \text{ (vehicles/hour)} \qquad (S1)$$

for simplicity, the effective green time-to-cycle length ratio $q$ is selected to be 0.5.

(2) when the speed limit of a road segment $45 < sl < 60$, it is defined as a highway:

$$C=(1,000+20 \times sl) \times l \text{ (vehicles/hour)} \qquad (S2)$$

(3) when the speed limit of a road segment $sl \geq 60$, it is defined as a freeway:

$$C=(1,700+10 \times sl) \times l \text{ (vehicles/hour)} \qquad (S3)$$

In Fig. S3, we show the distribution of road segment lengths. We find similar distributions in Bay Area and Boston Area, albeit the detected maximum length is larger in the Bay Area.

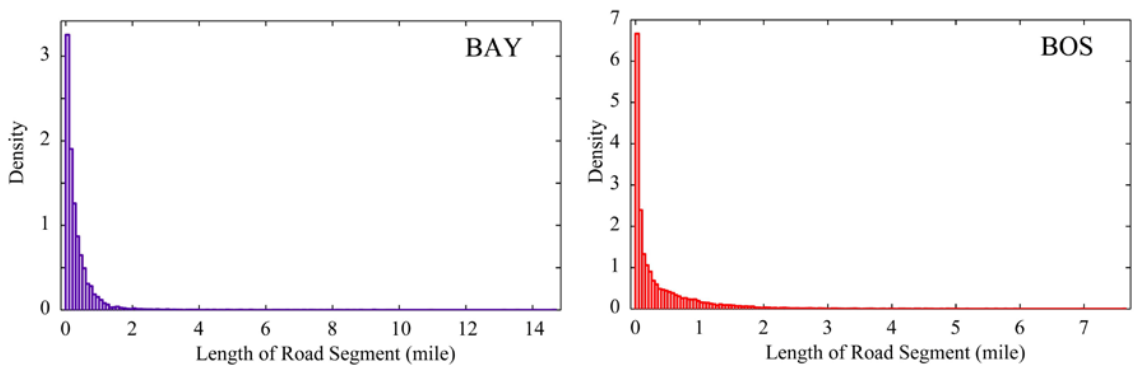

**Figure S3.** The distribution of road segment lengths.

## II. METHOD

### A. <u>Estimation of the Transient OD for Vehicle Users</u>

*1. Introduction:*

The Origin-Destination matrices (OD) provide information on flows of vehicles travelling from one specific geographical area to another and serve as one of the critical data inputs for transportation planning, design and operations (7). Currently OD is usually estimated from household interviews or incomplete traffic counts (8, 9). Traditional census and household interviews data fail to generate detailed and updated travel demands due to the high cost and low accuracy coupled with this method (8, 9). Road cameras and loop detectors can record the number of vehicles passing by, yet they are expensive to install and prone to errors and malfunctioning (8, 9), and consequently mostly limited to highways and freeways (8, 9). GPS data (10) collects location traces of probe vehicles at high resolutions (up to one Hz), yet they are not ubiquitous and fail to provide full OD information at a large scale. Furthermore, due to privacy issues they are often degraded on purpose (leading to down sampling of data), and thus insufficient as a standalone data source. Mobile phone data on the other hand, offer enormous amounts of location information, providing us with an opportunity to improve the estimation of the OD economically (11). An inherent advantage of mobile phone data comes from their wide availability. Because of the generic format of mobile phone data, any methodology relying on their analysis can easily be applied to other locations for which GIS data are also available, thus providing a unique framework pertinent to a variety of problems.

*2. Definition of trips and extraction of travel demands:*

The major challenge when estimating travel demands with mobile phone data is embedded in the sparse and irregular records (12), in which user displacements (consecutive different recorded locations) are usually observed between a long period (i.e. the first location is observed at 8:00am and next location is observed at 6:00pm). To more accurately extract users' travel demands between zones (mobile phone towers' service areas for the Bay Area and the census tracts for the Boston Area), we only record displacements occurring within a short time window. However, the time window we select must be long enough in order to ensure that enough travel demand information is extracted. In our modelling framework, we set the time window to one hour and define a *trip* as a displacement occurring within one hour in each time period (i.e. Morning Period, Noon & Afternoon Period, etc). Fig. S4 illustrates a mobile user's time and location records, using the presented approach; in this example two trips are detected.

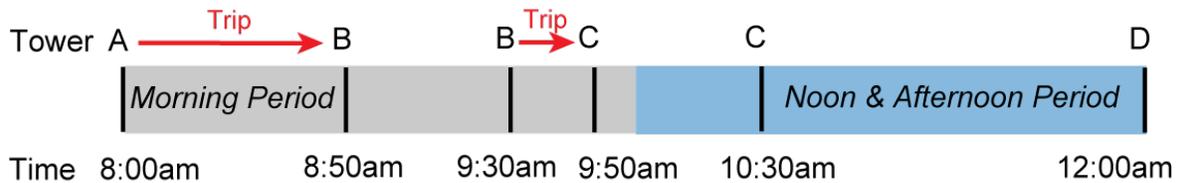

**Figure S4.** Illustration of trip definition from a mobile phone user's billing record. Black lines represent phone usage records, for each of them the time and the associated towers (A-D) routing the service are recorded. Changes of locations C->D are not defined as a trip, because they do not occur within a one-hour time window. Two trips are detected: from 8:00am tower A to 8:50am tower B and from 9:30am tower B to 9:50am tower C.

3. *Definition of transient OD:*

In the mobile phone data, a user's location information is lost when he/she does not use his/her phone. As Fig. S5 shows, a user is observed to move from zone B to zone C (he/she has calls or text messages in zone B and zone C), but his/her initial origin (O) and final destination (D) may actually be located in zone A and zone D. Thus, in such cases we lose a segment of the trip information (denoted by the dashed blue lines). Even if we only capture the transient origin and destination with the phones, this still allows us to capture a large portion of the road usage. Thus, we put forward the transient origin destination (*t*-OD) matrix, which requires only mobile phone data as input, to efficiently and economically capture the detailed travel demand information.

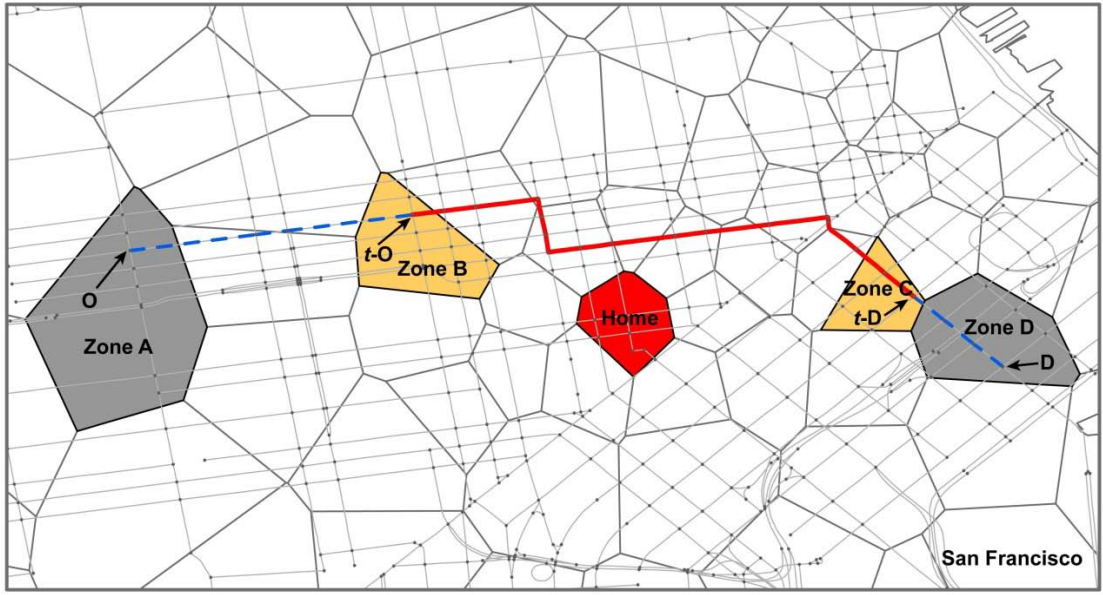

**Figure S5.** Illustration of a mobile phone user's OD, *t*-OD and home location**.** The road segments in the vicinity of San Francisco downtown are depicted by grey lines and the small black dots are the road intersections that lie in the zones (mobile phone towers' service areas). A driver drives from zone A (origin) to zone D (destination), however, he/she may only be detected by phone records at zone B (transient origin) and zone C (transient destination). The thick red line is the predicted route from the observed *t*-OD, whereas the dashed blue line represents the missing segment of the route. The driver's home location (driver source) is highlighted in red.

## 4. *Generation of travel demands independent of the frequency of phone activity:*

Obviously, users with more calls (text messages/web browsing) have more trips being extracted by the presented method. So one question arises: will this introduce bias to calculate the distribution of travel demands? To answer this question, we first measure the number of transactions (call/text message/web browsing) for the Bay Area and Boston Area users. As Fig. S6A shows, we find very similar distributions in the two areas. Thus, we use the same criterion to divide the mobile phone users into five groups, labelled I to V. The users in group I have less than 10 transactions, representing less than 5% of the user base. Group II, III, IV include the users with 10-500 transactions, 500-1,000 transactions and 1,000-2,000 transactions respectively, which overall represent ~90% of the selected users in the two areas. The mobile phone users in group V are extremely heavy users who have more than 2,000 transactions.

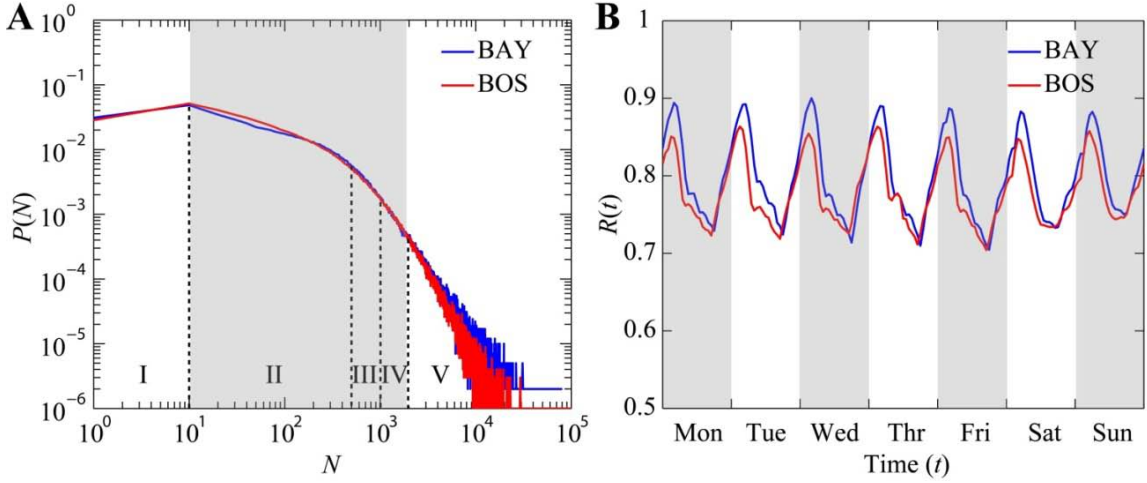

**Figure S6.** (**A**) The distribution of the number of transactions. $P(N)$ is the probability that a mobile phone user has $N$ transactions in three-week long observational period. Users are divided into five groups by the dashed lines and the users in group II, III and IV (the shaded area with grey colour) are used to extract trips between zones. (**B**) The hourly regularity $R(t)$ over a week-long period. It measures the probability when the user is found in his or her most visited location during the corresponding hour-long period.

We next count the number of trips $F_{ij}$ between zone $i$ and zone $j$ in a specific time period:

$$F_{ij} = \sum_{n=1}^{N} T_{ij}(n) \tag{S4}$$

where $N$ is the total number of selected users and $T_{ij}(n)$ is the total number of trips that user $n$ made between zone $i$ and zone $j$ in the observational period. The number of trips between zones $i$ and zone $j$ is then normalized by the total number of trips $\sum_{i,j} F_{ij}$ between all zones to obtain the distribution of travel demand $P_{ij}$:

$$P_{ij} = F_{ij} / \sum_{i,j} F_{ij} \tag{S5}$$

To test if $P_{ij}$ is sensitive to the selection of light or heavy users, we calculate $P_{ij}$ for users in group II, III, IV and V respectively (we do not use group I users, because they have too few locations recorded). We find that the $P_{ij}$ calculated from users in group II, III and IV are highly correlated (Pearson correlation coefficient *PCC*>0.93, Fig. S7), indicating that the distribution of travel demands is not sensitive to the selection of light or heavy users within a broad range. We find only a low *PCC* between users in groups II and V, consequently we do not take the small group of extremely heavy users (group V) into account. Thus we employ data from the user groups II, III and IV in our simulation.

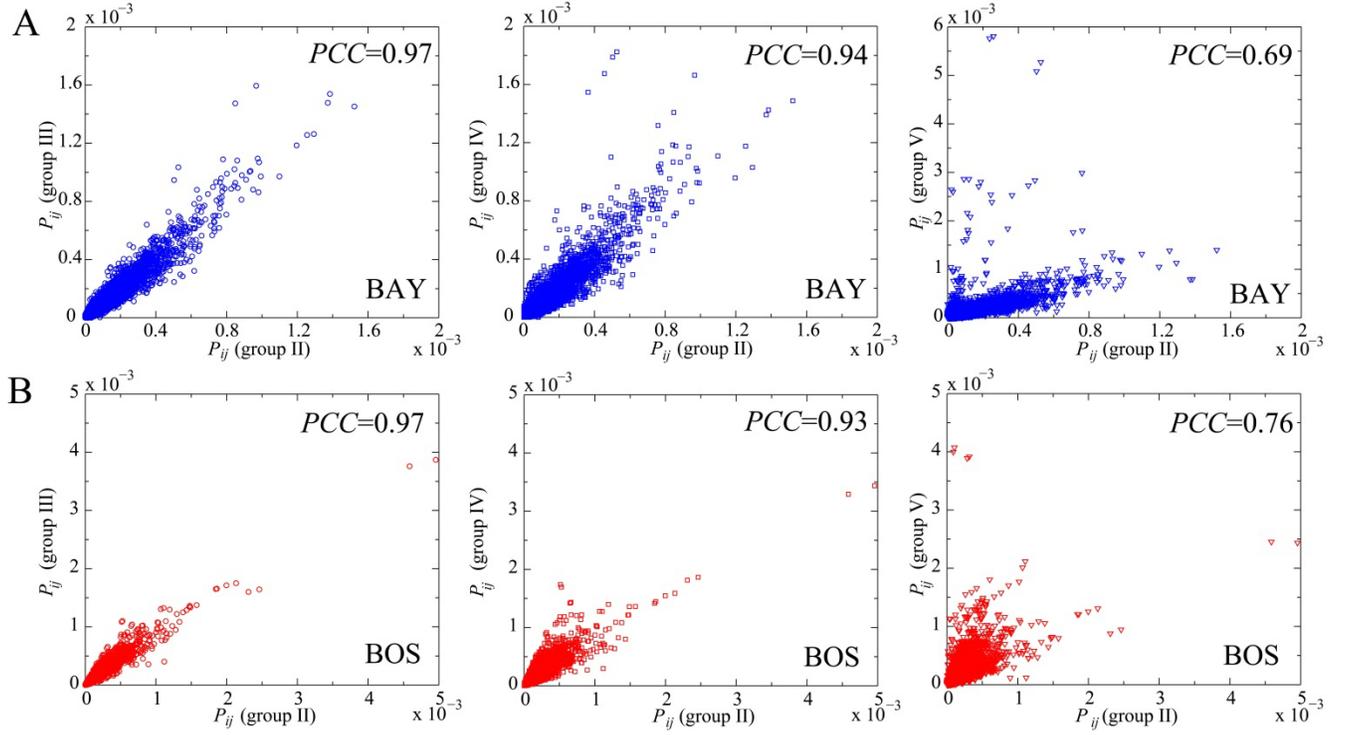

**Figure S7.** The distribution $P_{ij}$ of travel demands extracted from users in group II, III, IV and V. (**A**) In the Bay Area (BAY), $P_{ij}$ is extracted from group II, III, IV and V users respectively. The $P_{ij}$ extracted from users in groups II, III and IV are highly correlated, whereas a lower correlation is found between the $P_{ij}$ from group II and V users. To avoid the bias caused by these extremely active users, we employ users from group II, III and IV (91.5% of the selected 356,670 users) to extract the travel demand distribution. (**B**) Same as (A) but for the Boston Area (BOS) with 89.5% of the selected 683,001 users.

### 5. *Generating the vehicle based transient OD:*

One may note that the extracted distribution of travel demands did not take the population distribution into account. To avoid the bias caused by the unevenly distributed mobile phone user market share, we define the down-scale ratio ($M(i) < 1$) or the up-scale ratio ($M(i) \geq 1$) as follows:

$$M(i) = N_{pop}(i)/N_{user}(i) \tag{S6}$$

where $N_{pop}(i)$ and $N_{user}(i)$ are the population and the number of selected mobile phone users in zone $i$. The measured $M(i)$ distributions are shown in Fig. S8. For both areas, they are relatively broad, thus it is necessary to adjust the number of trips $F_{ij}$ by up-scaling or down-scaling the mobile phone users (Eq. S7).

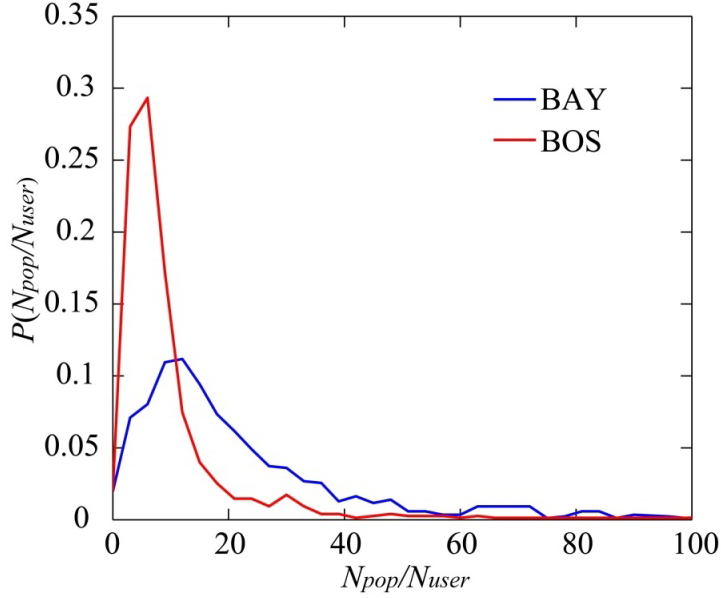

**Figure S8.** The blue curve corresponds to the distribution of up-scaling/down-scaling ratios $N_{pop}/N_{user}$ in the Bay Area (BAY) zones. The red curve corresponds to that in the Boston Area (BOS) zones. Note that in some regions the actual number of mobile phone users staying there may be larger than the number of residents registered by census.

After this process, the total number of trips generated by residents in a zone is proportional with its actual population:

$$F^{all}_{ij} = \sum_{n=1}^{N_k} T_{ij}(n) \times M(k) \tag{S7}$$

where $N_k$ is the total number of users in the $k^{th}$ zone and $T_{ij}(n)$ is the total number of trips that user $n$ made between zone $i$ and zone $j$ during the three weeks of study.

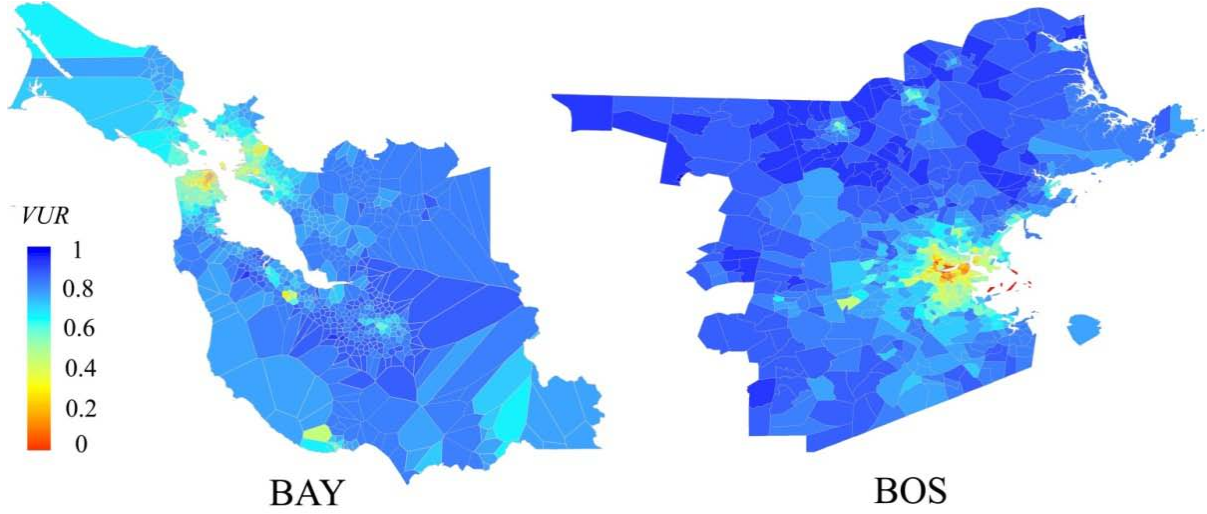

**Figure S9.** Vehicle usage rates by geographical area. Different colours represent different vehicle usage rates ($VUR$). Urban areas have lower $VUR$ than suburban areas, as can be noticed for San Francisco, a part of the east Bay and Santa Cruz, as well as for Boston.

People use different transportation modes throughout their trips. Possible transportation modes include car (drive alone), carpool, public transportation, bicycle and walk. We define a user is a vehicle user if he/she uses car to commute. We calculate the vehicle using rate ($VUR$) in a zone as follows:

$$VUR(i) = P_{car\ drive\ alone}(i) + P_{carpool}(i)/S \tag{S8}$$

where $P_{car\ driver\ alone}(i)$ and $P_{car\ pool}(i)$ are the probabilities that residents in zone $i$ drive alone or share a car. The average carpool size $S$ is 2.25 in California and 2.16 in Massachusetts (13). As shown in Fig. S9, $VUR$ is low in downtown and high in the suburb areas. Using the $VUR$ calculated for each zone, we randomly assign the transportation mode (vehicle or non-vehicle) to the users living in each zone. We then filter the trips that are not made by vehicles and calculate the total number of trips generated by vehicles $F^{vehicle}_{ij}$:

$$F^{vehicle}_{ij} = \sum_{n=1}^{N_k} T_{ij}(n) \times M(k) \tag{S9}$$

where user $n$ is a vehicle user, $N_k$ is the number of users in zone $k$.

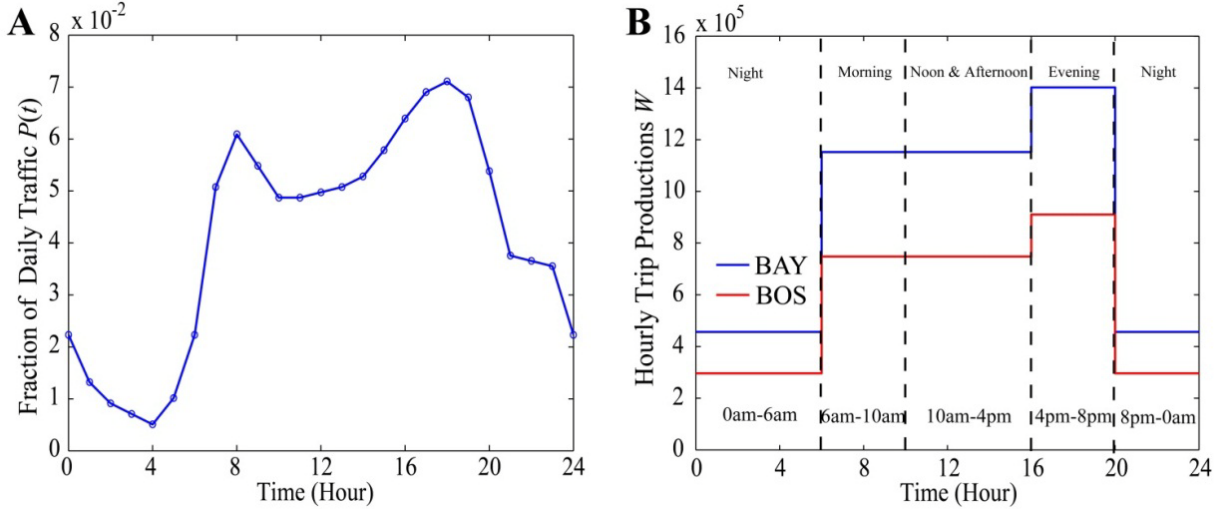

**Figure S10.** Distribution of daily traffic. (**A**) In each hour, the traffic contributed by vehicles represents a specific fraction of daily total traffic. (**B**) The average hourly total trip productions in the four time periods. For each time period, the hourly total trip productions are assigned as the average.

The average number of daily trips per person is about 4 in the US (14). This generates about 22 million trips in the Bay Area and 14 million trips in the Boston Area. Based on the daily distribution of traffic volume obtained from (15), we estimate the average hourly trip production $W$ in the four time periods (Fig. S10B). Next, we upscale the obtained distribution of travel demands with the hourly trip production $W$ for the entire population, thus finally defining the estimated $t$-OD.

$$t\text{-OD}_{ij} = W \times \frac{F^{vehicle}_{ij}}{\sum_{ij}^{A} F^{all}_{ij}} \tag{S10}$$

where $A$ is the number of zones. The following flow chart summarizes the methodology to calculate $t$-OD (Fig. S11).

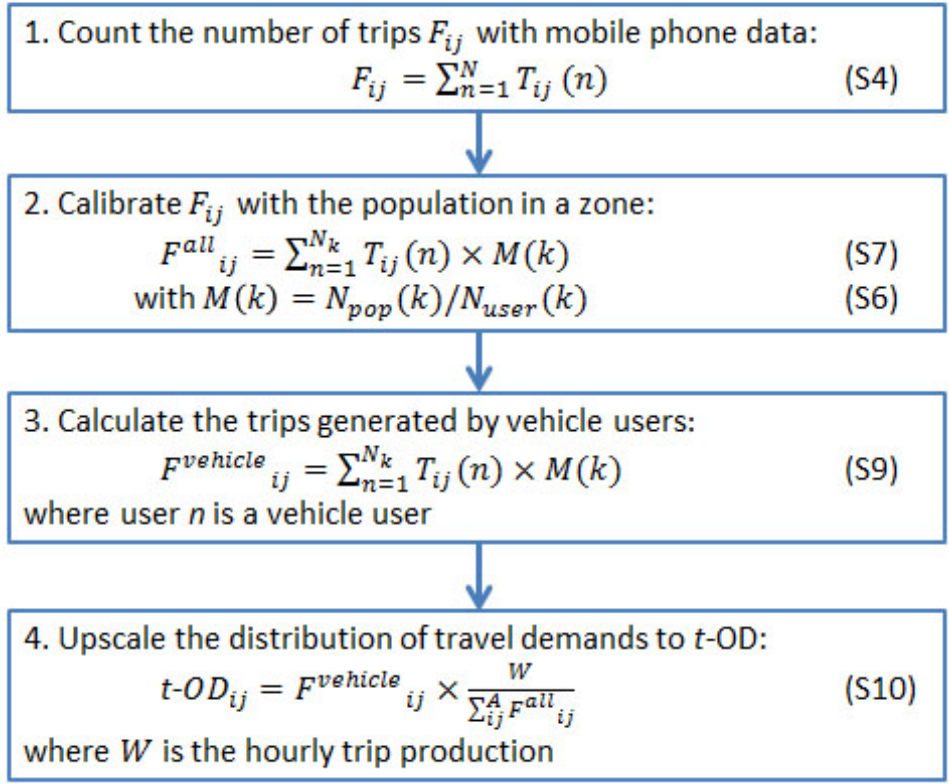

**Figure S11.** Flow chart for the calculation of *t*-OD.

**6.** *Converting zone based t-OD to intersection based t-OD:*

To assign trips to the road networks, we map each *t*-OD pair from zone based *t*-OD to intersection-based *t*-OD. We find the road intersections within a zone and randomly select one intersection to be the origin or destination in the intersection-based *t*-OD (Fig. S5). In very few cases no intersection is found in a zone. In such cases we assign a trip's origin or destination to a randomly chosen intersection in the nearest neighbouring zone. We generate four 11,096 × 11,096 intersection based *t*-OD from the four 892 × 892 zone based *t*-OD in the Bay Area (the Bay Area road network contains 11,096 intersections). For the Boston Area, we generate four 9,643 × 9,643 intersection based *t*-OD from the four 750 × 750 zone based *t*-OD (the Boston road network contains 9,643 intersections).

## B. Incremental Traffic Assignment

With the intersection based *t*-ODs calculated, we next assign the trips to the two road networks. The most fundamental method is provided by the classic Dijkstra algorithm, commonly used for routing in transportation networks (16). Dijkstra's algorithm is a graph search algorithm that solves the shortest path problem for a graph with nonnegative edge path costs (travel time in our case). With the Dijkstra algorithm, we can find the shortest path with minimum travel time between the origin and destination in a road network. However, the Dijkstra algorithm ignores the dynamical change of travel time in a road segment. Thus to incorporate the change of travel time, we apply the incremental traffic assignment (ITA) method (17) to assign the *t*-OD pairs to the road networks. In the ITA method, the original *t*-OD is first split into four sub *t*-ODs, which contain 40%, 30%, 20% and 10% of the original *t*-OD pairs respectively. These fractions are the commonly used values (18). The trips in the first sub *t*-OD are assigned using the free travel time $t_f$ along the routes computed by Dijkstra's algorithm. After the first assignment, the actual travel time $t_a$ in a road segment is assumed to follow the Bureau of Public Roads (BPR) function that widely used in civil engineering $t_a = t_f(1 + \alpha(VOC)^\beta)$, where commonly used values $\alpha = 0.15$ and $\beta = 4$ are selected (18). Next, the trips in the second sub *t*-OD are assigned using the updated travel time $t_a$ along the routes computed by Dijkstra's algorithm. Iteratively, we assign all of the trips in the four sub *t*-ODs. In the process of finding the path to minimize the travel time, we record the route for each pair of transient origin and transient destination.

The advantages of the ITA method consist of two aspects. First, it takes the dynamical change of travel time into account, mimicking the process of drivers selecting routes according to their knowledge of the traffic in a road network. Indeed, traffic flows predicted by the ITA method are a very good approximation of those predicted by the widely used User Equilibrium traffic assignment (UE) method (19). We find high correlations between the traffic flows predicted by the ITA method and the UE method in Fig. S12, which motivates the use of the ITA method for our work (it can be implemented easily without suffering from the computational complexity of UE solutions). Second, another advantage of the ITA method over the UE method is that by using the ITA method we can easily estimate the route of each OD pair, offering us the opportunity to study the road usage with respect to a road segment's driver sources (discussed in the main article).

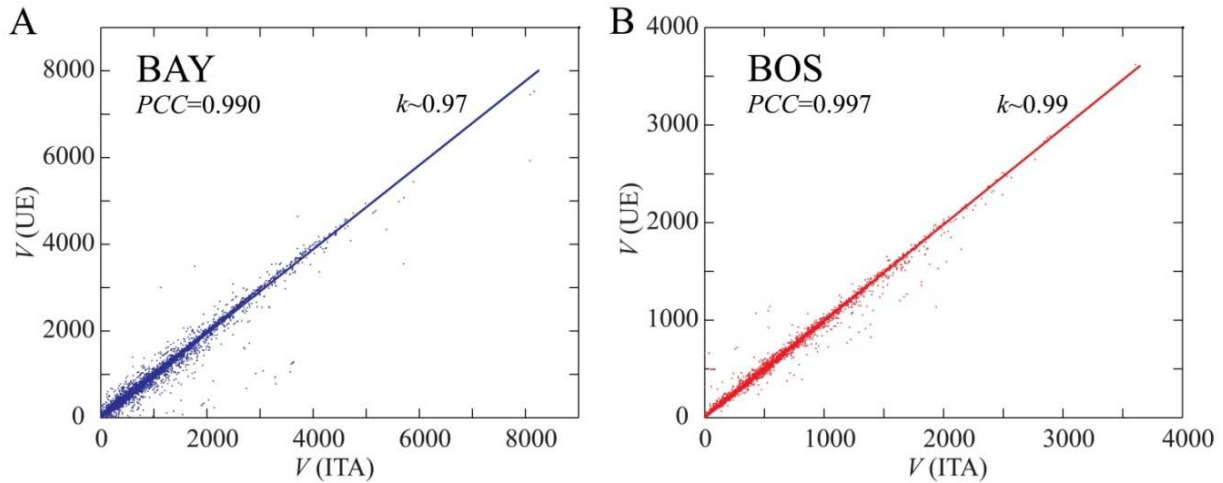

**Figure S12.** Validation of the ITA method. The x-axis represents the traffic flows (vehicles/hour) predicted by the ITA method and the y-axis represents that calculated by the UE method (UE function in TransCAD). The consistency of the results shows that the ITA method is a good approximation of the UE method. (**A**) shows the Bay Area (BAY). (**B**) shows the Boston Area (BOS).

## C. Estimation of Travel Time from GPS Probe Data

In order to validate the results from the previous sections, an independent data set is needed in order to compare the corresponding estimates with these independent measurements. Probe vehicle data based on GPS receivers has enjoyed a widespread use in transportation. However it must be said that it will not be possible in the near future to use GPS probe data to calculate traffic volumes in whole urban road networks. This is because the amount of probe data is still too low to be used for inference of traffic volumes. Probe data has successfully been used to compute travel times and speeds along freeways and arterials (20). Thus, the validation process used to assess the accuracy of our method will rely on travel time and speed as a proxy, which we can infer from probe data provided by taxicabs and commercial vehicle companies.

This data show unique advantages for tracking a fleet of vehicles and routing and navigation. The receivers are usually attached to a car or a truck (referred to as a probe vehicle), and they relay information to a base station using the data channels of the cell phone networks. A datum provided by probe vehicles includes an identifier of the vehicle, a GPS position and a timestamp. In order to reduce power consumption and transmission costs, the probe vehicles do not continuously report their location to the base station. Instead they relay their position either at fixed times (every second to every minute), or at some landmark positions (a concept patented by Nokia under the term Virtual Trip Time) (21). This data type is very popular, especially amongst transportation companies for tracking purposes, but presents unique challenges for estimating traffic flows patterns:

(1) The precise location of the vehicle is known with some error, due to GPS observation noise.
(2) The path of the vehicle between two consecutive observations can be significantly long, and is usually unobserved.

The approach used in this work is to reconstruct the trajectories of the vehicles as accurately as possible, using machine learning techniques. From these trajectories, only sample points are observed, between which the travel time is known. This information (travel time, reconstructed trajectory) is then passed on to a second learning algorithm that learns travel times on every road link. This process is repeated for every day of the week and every 15 minutes of a day to calculate a weekly historical estimate of the traffic. We briefly describe the mapping algorithm below and then introduce the travel time learning algorithm (Fig. S13).

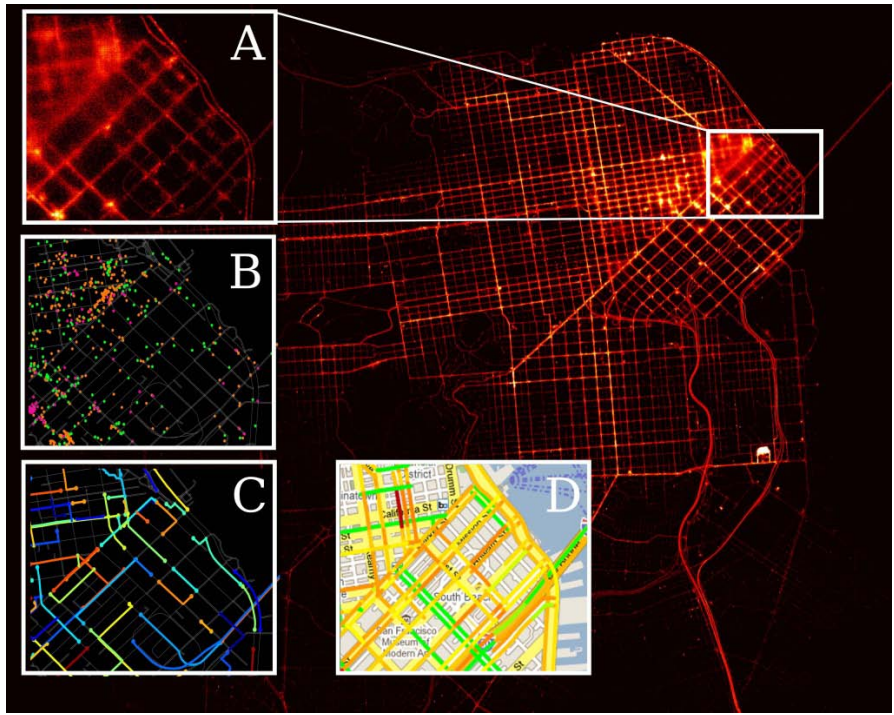

**Figure S13.** Estimation of travel times using probe vehicle data. In the background, the density map of probe data around San Francisco is shown. The maximum density (in white) corresponds to 7.2 GPS observations per hour and per square meter. (**A**) focuses on the Embarcadero neighborhood. (**B**) shows the GPS observations (sent every minute) collected from three vehicles in that area between 8am and 10am. The trajectory of each vehicle is reconstructed from the sequence of GPS points using the Path Inference algorithm. (**C**) presents a few trajectory segments between two consecutive GPS point. The EM algorithm then infers the travel times on each road link, by learning from these time-stamped segments. (**D**) shows a typical output of the travel time algorithm, at 8am on a Monday Morning.

*Map Matching Algorithm*

The GPS error is assumed to follow a (nearly Gaussian) dispersion model. Meanwhile, the driver's behaviour on the road is assumed to follow a model that indicates the preferences of the driver between one path and another. Our framework can be decomposed into the following steps:

Map matching: each GPS measurement from the input is projected onto a set of candidate states on the road network. The vehicle is assumed to have been in either of these candidate states when the GPS observation was made.

Path discovery: a number of potential paths are computed between pairs of candidate states on the road network. The vehicle is assumed to have followed one of these paths when it travelled from the previous observation to the next

Filtering: probabilities are assigned to the paths and the states using a model of the driver's preferences and of the GPS dispersion. These probabilities are computed using a dynamic programming approach, using a probabilistic structure called a Conditional Random Field. Using the Viterbi algorithm, the most likely trajectory is obtained. At the output of the filter, we obtain reconstructed trajectories, along with time stamped waypoints. This dataset is then used to computing historical travel time estimates.

*Expectation Maximization Algorithm*

Each segment of the trajectory between two GPS points is referred to as an observation. An observation consists of a start time, an end time and a path on the road network. This path may span multiple road links, and starts and ends at some offset within some links. The observations are grouped into 15 minute time intervals and sent to a traffic estimation engine, which runs the learning algorithm described next and returns probability distributions of travel times for each link. The goal of the traffic estimation algorithm is to infer how congested the links are in a road network, given periodic GPS readings from vehicles moving through the network. An additional difficulty in estimating the travel time distributions is the lack of travel times for the individual links. Instead, each observation only specifies the total travel time for an entire list of links travelled. To solve this problem, we use an iterative expectation maximization (EM) algorithm. The central idea of the algorithm is to randomly partition the total travel time among links for each observation, and then weigh the partitions by their likelihood according to the current estimate of travel time distributions. Next, given the weighted travel time samples produced for each link, we update the travel time distribution parameters for the link to maximize the likelihood of these weighted samples. By iteratively repeating this process, the algorithm converges to a set of travel time distribution parameters that fit the data well. The sample generation stage is called the expectation (E) step, and the parameter update stage is called the maximization (M) step. This procedure rapidly and reliable converges to some estimated travel times for every road of the network.

## D. <u>Validation</u>

Due to the lack of reliable traffic flow data at a global scale (due to the insufficient volume of probe data), we compare for each road segment the predicted travel time with the average travel time calculated from the probe vehicle GPS data (the data is mostly obtained from Taxi fleets). According to the BPR function, the travel time of a road segment is decided by its traffic flow. A road segment's travel time increases with the increase of its traffic flow. Hence, obtaining the travel time from GPS probe data can be an indirect way to validate our results on the distribution of traffic flow. For 68% of the road segments in the Bay Area road network (16,594), the probe vehicle GPS data record the average travel time in each 15 minute interval of the one week observational period. Using this data, we calculate the average travel time for each road segment in the four time periods considered for this work (Morning, Noon & Afternoon, Evening and Night). We find that the predicted travel time from the *t*-OD has a good linear relation $T_{\text{prediction}} = kT_{\text{probe vehicle}}$ with the average travel time estimated from the probe vehicle GPS data (the coefficient of determination $R^2 > 0.9$ for all time periods). The Pearson correlation coefficients (*PCC*) are larger than 0.95 for all time periods (Fig. S14). The slope $k$ is about 0.75 in the daytime, which may be caused by the relatively frequent waiting or speed deceleration when drivers wait at traffic lights (we did not consider traffic signals in the presented modelling framework). The slope is about 1 in the Night period, indicating the high vehicle speeds during this period. Taken together we find a high correspondence between our predicted result and the GPS probe data estimation, demonstrating the strength of the presented methodology. Furthermore, elements such as more accurate information about road capacity, free travel time and parameters for the BPR function and traffic signals can be integrated into our fundamental modelling framework to enrich future predictions.

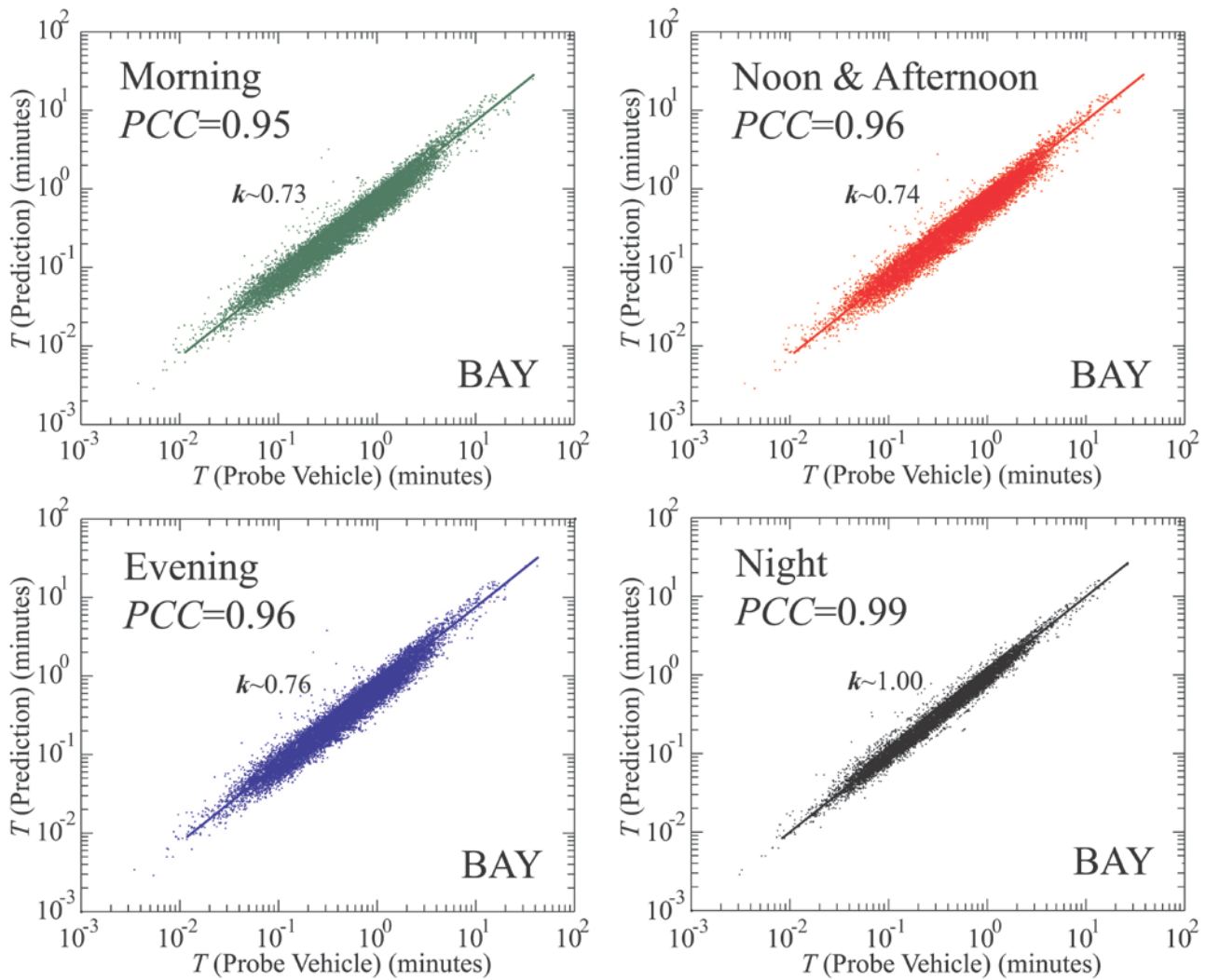

**Figure S14.** The predicted travel time is validated by the travel time estimated from the probe vehicle GPS data. Because traffic flow data is not available on arterial roads, the only available comparison variable to assess the validity of the method is travel time (which can be measured directly from probe data). To this day, this is the only feasible method to perform this comparison at a global scale and represents the latest state of the art.

## III. RESULTS

### A. Supplementary Results

#### 1. The road segment's degree is lowly correlated with traditional measures:

As Fig. S15 shows, although relatively large Pearson correlation coefficient $PCC$=0.65 (BAY) and $PCC$=0.60 (BOS) are measured, road segments with similar traffic flow can still have large difference in their $K_{\text{road}}$. We also find road segments with similar $VOC$ can have very different $K_{\text{road}}$ ($PCC$=0.46 and $PCC$=0.37 for the Bay Area and the Boston Area respectively). This result indicates that for road segments with similar condition of congestion, the diversity of their driver sources may be very different. The betweenness centrality $b_c$ of a road determines its ability to provide a path between separated regions of the network. We find $b_c$ also has low correlations with $K_{\text{road}}$ (Fig. S15C and F).

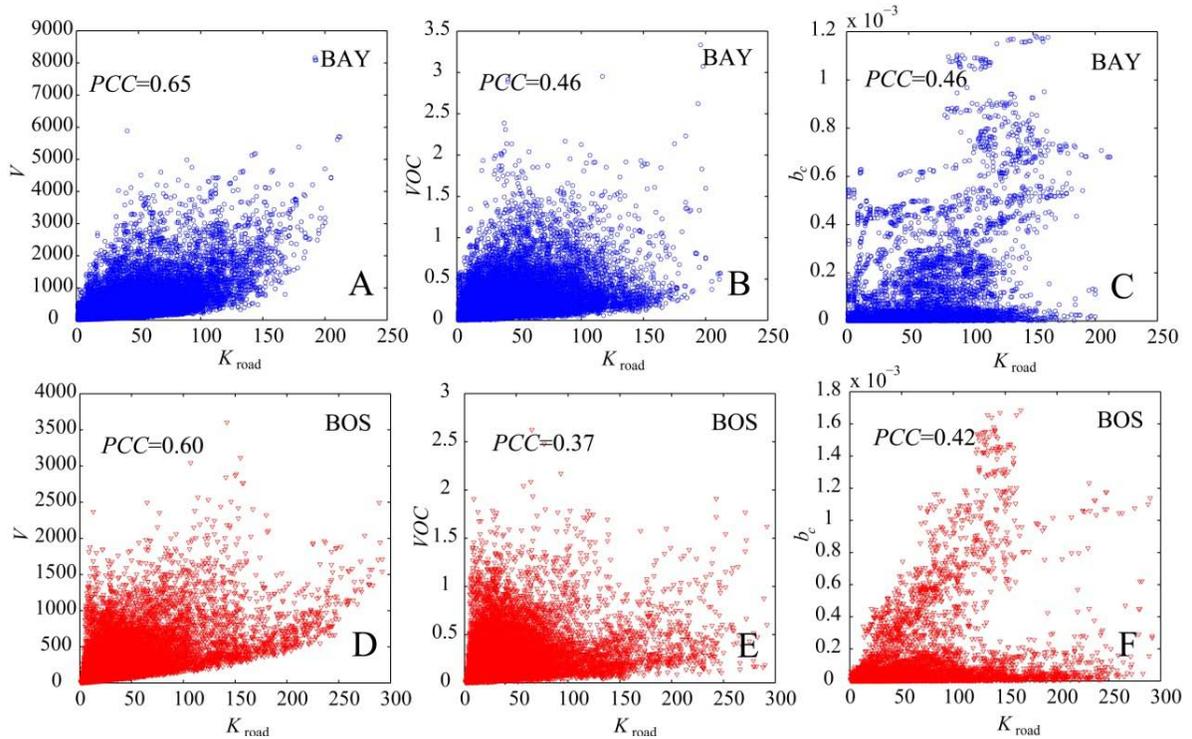

**Figure S15.** Road segment's degree $K_{\text{road}}$ has low correlations with its traffic flow $V$, $VOC$ and betweenness centrality $b_c$. (**A**) Pearson correlation coefficient ($PCC$) between $V$ and $K_{\text{road}}$ in the Bay Area. (**B**) $PCC$ between $VOC$ and $K_{\text{road}}$ in the Bay Area. (**C**) $PCC$ between $b_c$ and $K_{\text{road}}$ in the Bay Area. (**D**), (**E**), (**F**) Same as (A), (B), (C) respectively but for the Boston Area.

## 2. Grouping the road segments according to their $b_c$ and $K_{road}$:

Fig. S16 shows the betweenness centrality $b_c$ and the degree $K_{road}$ of road segments. Road segments are grouped and depicted in different colors.

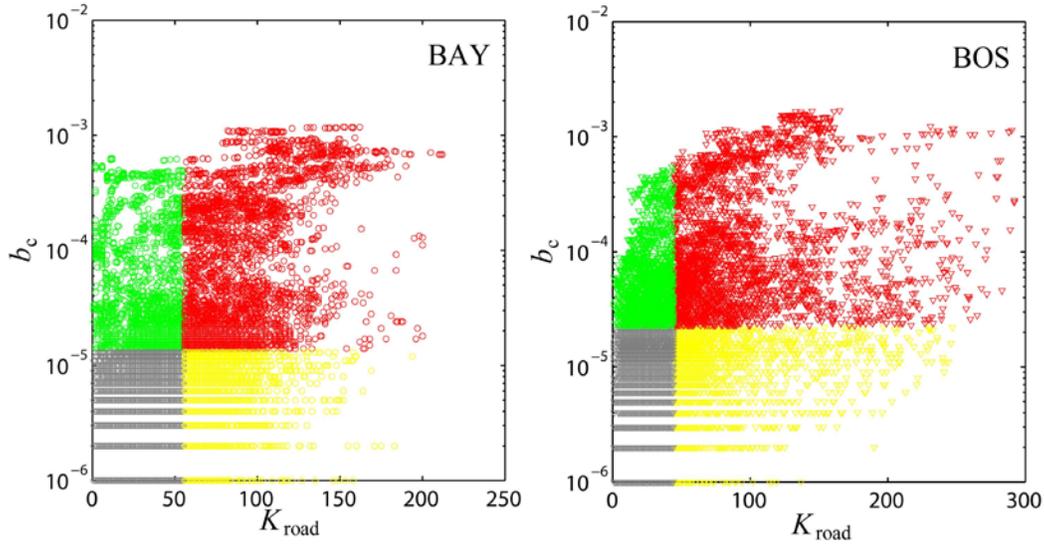

**Figure S16.** Types of roads defined by $b_c$ and $K_{road}$. The road segments are grouped by their betweenness centrality $b_c$ and degree $K_{road}$. The red symbols represent the roads with the largest 25% of $b_c$ and $K_{road}$. The green symbols represent those with the largest 25% of $b_c$ and the smallest 75% of $K_{road}$. The yellow symbols are those with the smallest 75% of $b_c$ and the largest 25% $K_{road}$. The road segments depicted in grey have the smallest 75% of $b_c$ and $K_{road}$.

## 3. The total additional travel time $T_e$ in driver sources:

Fig. S17 shows the total additional travel time $T_e$ of the driver sources. Due to the heterogeneity of road usage, $T_e$ is very unevenly distributed in space in the two metropolitan areas, enabling us to easily locate the driver sources with high $T_e$. For the Bay Area, the top 1.5% driver sources (12 sources) with the largest $T_e$ are selected. In the case study for the Boston Area, we select 15 driver sources (top 2%) with highest $T_e$. This selection makes sure that for a similar local trip reduction *f*, the global trip reduction *m* is same as that of the Bay Area.

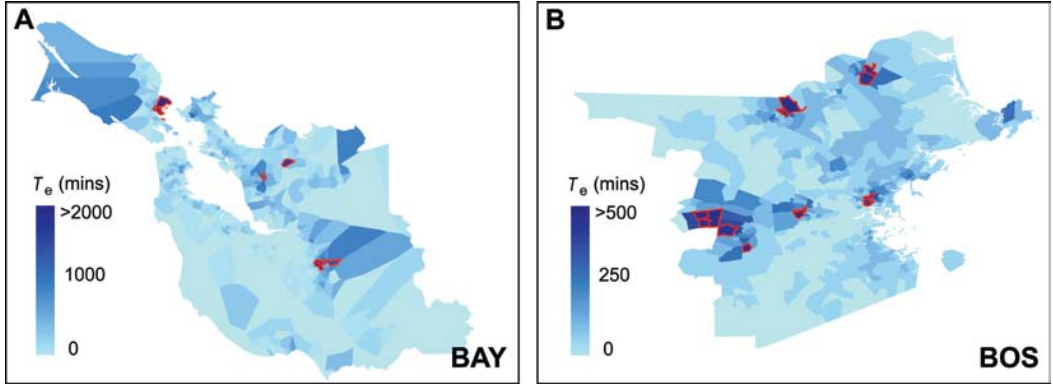

**Figure S17.** (**A**) The total additional travel time $T_e$ for each Bay Area driver source. The red polygons locate the pinpointed driver sources with $T_e > 1,355$ minutes. Thus the drivers suffering from heavy traffic congestion are located. (**B**) Same as (A) but for the Boston Area. The red polygons locate the targeted 15 driver sources (top 2%) with a total of more than 400 minutes additional travel time in one hour of the morning commute.

To address the underlying reasons for the high efficiency of the selective strategy (Fig. 4*B*), we measure the average traffic flow reduction $\delta V$ for road segments with different levels of $VOC$. As Fig. S18 shows, the red, green and blue curves correspond to the road segments with $VOC > 1$ (High $VOC$), $0.5 < VOC \leq 1$ (Middle $VOC$) and $VOC \leq 0.5$ (Low $VOC$) respectively. We find that for high $VOC$ road segments, $\delta V$ is much larger in a selective strategy for both Bay Area and Boston Area, indicating that the selective strategy can more efficiently decrease the traffic flows in the congested road segments.

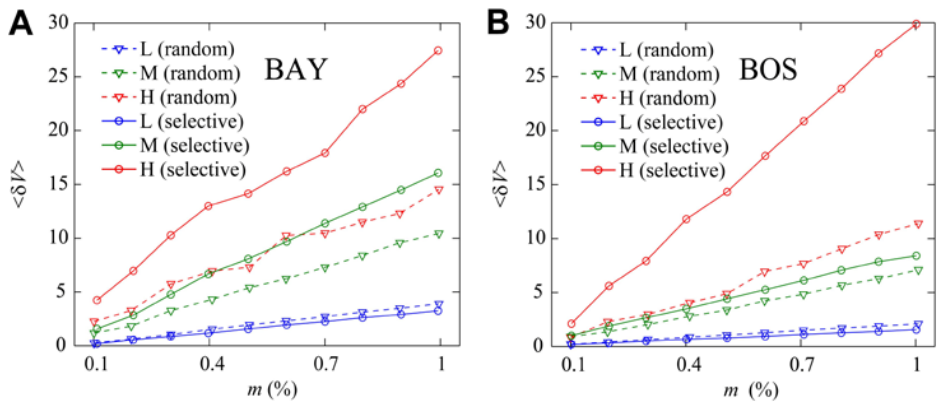

**Figure S18.** (**A**) The average traffic flow reduction $<\delta V>$ over road segments with different $VOC$ in the Bay Area. Red, green and blue symbols correspond to road segments with $VOC > 1$, $0.5 < VOC \leq 1$ and $VOC \leq 0.5$ respectively. (**B**) Same as (A) but for the Boston Area.

## 4. The results for other time periods:

Fig. S19 are counterpart figures for Fig. 1 and Fig. 2. It shows the corresponding results in the other three periods (Noon & Afternoon, Evening and Night). We find that the results for the three daytime periods show high similarities, whereas the results in the Night period are different due to minor road usage. These results indicate that using our modelling framework, we can capture the road usage pattern dynamically.

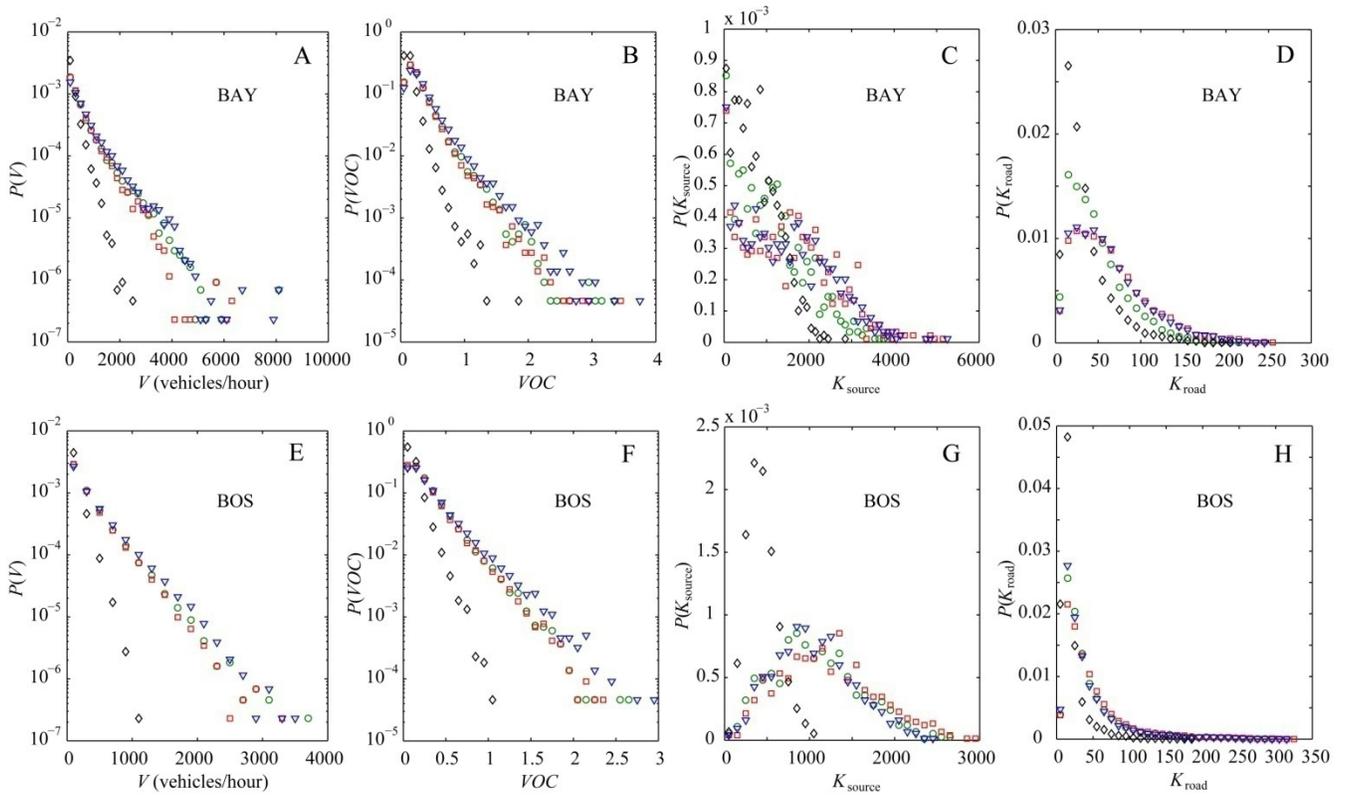

**Figure S19.** Green circles represent the results in Morning period, red squares represent the results in Noon & Afternoon period, blue triangulations represent the results in Evening period and black diamonds represent the results in Night period. (**A**) Distribution of the Bay Area one-hour traffic flow in the four time periods. The one-hour traffic flow in the Night period is much smaller than that found in the daytime periods. (**B**) Distribution of the Bay Area $VOC$ in the four time periods. (**C**) Degree distributions of the Bay Area driver sources in the four time periods. (**D**) Degree distributions of the Bay Area road segments in the four time periods. (**E**), (**F**), (**G**), (**H**) are same as (A), (B), (C), (D) respectively but for the Boston Area.

## 5. *The distance from road segment to its MDS:*

We measure the distance *d* from each road segment to its major driver sources. We find that *d* can be well approximated by a log-normal distribution $P(d) \sim e^{-(\ln(d)-\mu)^2/2\sigma^2}/(\sqrt{2\pi}\sigma d)$. As Fig. S20 shows, the distance *d* centers around 4km and 7km for Bay Area and Boston Area respectively, indicating that the MDS are geographically nearby the corresponding road segment. However, there exist MDS that are far away from the road segment (>50km). The prediction of these specific MDS is beyond a traditional distance decaying function, and this is the power of our modeling framework in capturing the urban travel demand.

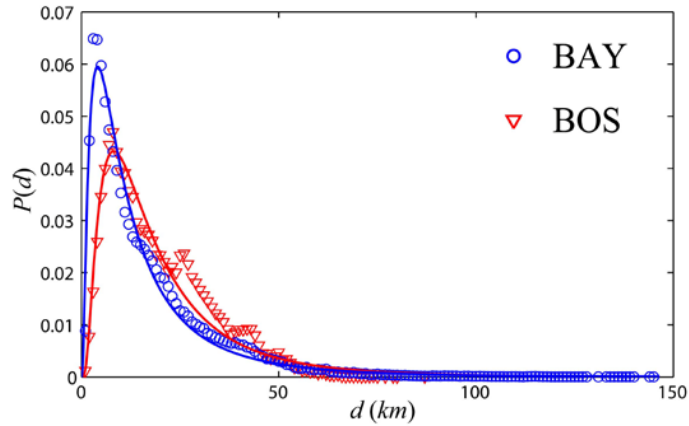

**Figure S20.** The distribution of the distance *d* from each road segment to its MDS. The blue circles represent the result for Bay Area and the red triangles represent the result for Boston Area. The distance *d* from each road segment to its MDS can be well approximated by a log-normal distribution $P(d) \sim e^{-(\ln(d)-\mu)^2/2\sigma^2}/(\sqrt{2\pi}\sigma d)$ with $\mu = 2.40$ (2.76), $\sigma = 0.99$ (0.81), $R^2 = 0.98$ (0.96) for Bay Area (Boston Area).

## B. Statistical Analysis

The purpose of this section is to support our findings with rigorous goodness-of-fit analysis. We evaluate goodness-of-fit statistics for parametric models in the paper by calculating the sum of squares due to error (SSE), the $R^2$ and the root mean squared error (RMSE).

| Area | $p_A$ | $p_H$ | $\beta_H$ | $c_A$ | $\alpha_A$ | $R^2$ | SSE | RMSE |
|---|---|---|---|---|---|---|---|---|
| BAY | 0.876 | 0.124 | 0.00026 | 4.625e-008 | 2.43 | 0.991 | 5.38e+006 | 227.4 |
| BOS | 0.942 | 0.058 | 0.00046 | 2.581e-005 | 1.86 | 0.996 | 2.76e+006 | 135.8 |

**Table S2.** The distribution of betweenness centrality: $P(b_c) = p_H \beta_H e^{-b_c/\beta_H} + p_A c_A b_c^{-\alpha_A}$. $p_A$ is the fraction of arterial roads, $p_H$ is the fraction of highways, $\beta_H$ is the average highway betweenness centrality, $c_A$ and $\alpha_A$ are estimated by the Matlab fitting toolbox.

| Area | $p_A$ | $p_H$ | $v_A$ | $v_H$ | $R^2$ | SSE | RMSE |
|---|---|---|---|---|---|---|---|
| BAY | 0.876 | 0.124 | 373 | 1493 | 0.999 | 4.126e-009 | 1.193e-005 |
| BOS | 0.942 | 0.058 | 236 | 689 | 0.997 | 2.508e-008 | 3.959e-005 |

**Table S3.** The distribution of one-hour traffic flow: $P(V) = p_A v_A e^{-V/v_A} + p_H v_H e^{-V/v_H}$. $p_A$ is the fraction of arterial roads, $p_H$ is the fraction of highways, $v_A$ is average traffic flow for arterial roads, $v_H$ is the average traffic flow for highways.

| Area | $\gamma$ | $R^2$ | SSE | RMSE |
|---|---|---|---|---|
| BAY | 0.28 | 0.983 | 0.2199 | 0.08861 |
| BOS | 0.28 | 0.982 | 0.1769 | 0.08769 |

**Table S4.** The distribution followed by $VOC$: $P(VOC) = \gamma e^{-VOC/\gamma}$, $\gamma$ is the mean of $VOC$.

| Area | $\tau$ | $R^2$ | SSE | RMSE |
|---|---|---|---|---|
| BAY | 204.2 | 0.972 | 2.362e-006 | 0.000198 |
| BOS | 113.2 | 0.735 | 6.038e-005 | 0.001374 |

**Table S5.** The distribution followed by total additional travel time: $P(T_e) = \tau e^{-T_e/\tau}$, $\tau$ is the mean of $T_e$.

| Area | $\mu_{source}$ | $\sigma_{source}$ | $R^2$ | SSE | RMSE |
|---|---|---|---|---|---|
| BAY | 1035.9 | 792.2 | 0.785 | 2.847e-007 | 8.893e-005 |
| BOS | 1017.7 | 512.3 | 0.914 | 1.540e-007 | 7.849e-005 |

**Table S6.** The statistical fits of driver source's degree: $P(K_{source}) = e^{-(K_{source}-\mu_{source})^2/2\sigma_{source}^2}/(\sqrt{2\pi}\sigma_{source})$, where $\mu_{source}$ is the mean of $K_{source}$ and $\sigma_{source}^2$ is the variance.

| Area | $\mu_{road}$ | $\sigma_{road}$ | $R^2$ | SSE | RMSE |
|---|---|---|---|---|---|
| BAY | 3.713 | 0.821 | 0.978 | 9.013e-006 | 0.0008326 |
| BOS | 3.359 | 0.724 | 0.989 | 1.271e-005 | 0.0006737 |

**Table S7.** The statistical fits of road segment's degree:
$P(K_{road}) = e^{-(\ln(K_{road})-\mu_{road})^2/2\sigma_{road}^2}/(\sqrt{2\pi}\sigma_{road} K_{road})$

| Strategy | $k$ | $b$ ($b\sim 0$) | $R^2$ | SSE | RMSE |
|---|---|---|---|---|---|
| BAY (Random) | 6.931e+005 | -0.0020 | 0.9019 | 4.349e+006 | 737.3 |
| BAY (Selective) | 2.261e+006 | -0.0010 | 0.9820 | 7.753e+006 | 984.4 |
| BOS (Random) | 2.300e+005 | 0.0015 | 0.9818 | 8.299e+004 | 101.9 |
| BOS (Selective) | 1.181e+006 | -0.0001 | 0.9966 | 4.023e+005 | 224.3 |

**Table S8.** The total additional travel time reduction $\delta T$ with the trip reduction percentage $m$ in the cases of selective and random strategies: $\delta T = k(m - b)$.

| Periods | $k$ | $R^2$ | SSE | RMSE |
|---|---|---|---|---|
| Morning | 0.7277 | 0.9093 | 1701 | 0.3204 |
| Noon & Afternoon | 0.7416 | 0.9200 | 1496 | 0.3005 |
| Evening | 0.7567 | 0.9178 | 1541 | 0.3049 |
| Night | 0.9951 | 0.9724 | 516.3 | 0.1765 |

**Table S9.** The validation of predicted travel time by the estimated travel time from probe vehicles' GPS data: $T_{prediction} = kT_{probe\ vehicle}$.